\documentclass[apj]{emulateapj}
\usepackage{apjfonts} 
\usepackage{lscape}

\usepackage{graphicx}

\newcommand{\bc}{\begin{center}}
\newcommand{\ec}{\end{center}}
\newcommand{\cf}{\ifmmode C_f\else $C_f$\fi}

\submitted{February 6, 2010}

\slugcomment{Accepted for publication in the {\it Astronomical Journal}.}
\shortauthors{Chen et al.}
\shorttitle{Outflow traced by Na D absorption}
\begin{document}
\title{Absorption-line probes of the prevalence and properties of outflows\\ 
in present-day star-forming galaxies}
\author{
Yan-Mei Chen\altaffilmark{1},
Christy A. Tremonti\altaffilmark{2},
Timothy M. Heckman\altaffilmark{3},
Guinevere Kauffmann\altaffilmark{1},
Benjamin J. Weiner\altaffilmark{4},
Jarle Brinchmann\altaffilmark{5},
Jing Wang\altaffilmark{1,6}
}
\altaffiltext{1}{
Max--Planck--Institut f\"ur Astrophysik,
    Karl--Schwarzschild--Str. 1, D-85748 Garching, Germany
\email{chenym@mpa-garching.mpg.de}
}
\altaffiltext{2}{
Department of Astronomy, University of Wisconsin-Madison, 1150 University Ave, Madison, WI 53706, USA
}
\altaffiltext{3}{
Department of Physics and Astronomy, The Johns Hopkins University, 3400 North Charles Street, Baltimore, MD 21218, USA
}
\altaffiltext{4}{
Steward Observatory, 933 N. Cherry St., University of Arizona, Tucson, AZ 85721, USA
}
\altaffiltext{5}{
Leiden Observatory, Leiden University, 2300 RA, Leiden, The Netherlands
}
\altaffiltext{6}{
Center for Astrophysics, University of Science and Technology of China, 230026 Hefei, China
}

\begin{abstract}
  We analyze star forming galaxies drawn from SDSS DR7 to show how the
  interstellar medium (ISM) Na I $\lambda\lambda$5890, 5896 (Na D)
  absorption lines depend on galaxy physical properties, and to look for
  evidence of galactic winds. We combine the spectra of galaxies with
  similar geometry/physical parameters to create composite spectra
  with signal-to-noise $\sim300$. The stellar continuum is modeled
  using stellar population synthesis models, and the
  continuum-normalized spectrum is fit with two Na~I absorption
  components. We find that: (1) ISM Na D absorption lines with
  equivalent widths EW $>$ 0.8~\AA\ are only prevalent in disk
  galaxies with specific properties -- large extinction ($\rm A_V$),
  high star formation rates (SFR), high star formation rate per unit 
  area ($\Sigma_{\rm SFR}$), or high stellar mass ($M_*$).  (2) the ISM 
  Na D absorption lines can be separated into two components: a quiescent 
  disk-like component at the galaxy systemic velocity and an outflow 
  component; (3) the disk-like component is much stronger in the edge-on systems, and
  the outflow component covers a wide angle but is stronger within
  $60\,^{\circ}$ of the disk rotation axis; (4) the EW and covering factor of the 
  disk component correlate strongly with dust attenuation, highlighting the
  importance that dust shielding may play in the survival of Na~I. (5) The EW of the 
  outflow component depends primarily on $\Sigma_{\rm SFR}$ and secondarily on $A_V$; 
  (6) the outflow velocity varies from $\sim120$ to 160~km~s$^{-1}$ but shows little 
  hint of a correlation with galaxy physical properties over the modest dynamic range 
  that our sample probes (1.2 dex in log$~\Sigma_{\rm SFR}$ and 1 dex in $\log M_*$). 

\end{abstract}

\keywords{
galaxies : evolution ---
galaxies : star formation 
}

\section{Introduction}
\label{sec:intro}
Galactic-scale gaseous outflows (`galactic winds') are known to be
ubiquitous in very actively star forming galaxies at all cosmic epochs
\citep{heckman90, pettini00, shapley03, menard09, weiner09}.  
Galactic winds play an vital role in the evolution of galaxies and the
intergalactic medium (IGM).  The ``baryon deficit" in the Galaxy
\citep[e.g.,][]{silk03} indicates gas removal by outflows during past
active episodes.  The mass-metallicity and effective yield relations
observed in local galaxies suggest that galactic winds transport
highly metal-enriched gas out of galaxies and into the IGM
\citep[e.g.,][]{garnett02,tremonti04,dalcanton07}.  
The $\Lambda$CDM model over-predicts the galaxy luminosity function at
both the low and high luminosity ends if a constant mass-to-light ratio is
assumed. The most natural way to reconcile the observations with
theory is to invoke feedback processes, including supernova, stellar
winds \citep{Cole00, Benson03, stringer08, Oppenheimer09} and AGN activity
\citep{Silk98, Hopkins06}.  However, it remains unclear which kind of
feedback processes dominate as a function of luminosity and comic
epoch.




In actively star forming galaxies, galactic winds are driven by the mechanical 
energy and momentum imparted by stellar winds and supernovae 
\citep{chevalier85, heckman90}.  Young star clusters create over-pressured 
bubbles of coronal phase gas which expand and sweep up shells of
ambient ISM until they `blow-out' of the disk into the halo.
The collective action of multiple superbubbles drives a weakly collimated
bi-polar outflow consisting of hot gas and cool entrained clouds.
Radiation pressure is also likely to play a role in accelerating cool 
   dusty material \citep[e.g.,][]{murray05}.
In this paper we use the term `galactic wind' to
describe such outflows without regard for their eventual fate ---
the gas may escape the halo potential well, or be recycled
back into the disk in a process sometimes referred to as 
`galactic fountain' activity \citep{shapiro76, bregman80, kahn81}.

In the last two decades, there have been many attempts to directly
observe galactic winds in galaxies at $z=0- 3$ \citep{veilleux05}. 
In the local universe,
outflows can be detected via X-rays which trace hot gas
\citep[e.g.,][]{dahlem98,martin02,strickland04}, optical nebular emission
lines produced by warm gas \citep[e.g.,][]{lehnert96} and ISM
absorption lines (e.g., Na I, K I) from cold gas
\citep[e.g.,][]{heckman02}. Blue-shifted Na D absorption from the
entrained cool gas is frequently detected in IR-bright starburst
galaxies \citep{heckman00}, LIRGs and ULIRGs \citep{rupke02, rupke05a,
  rupke05b, martin05, martin06, martin09}, and it is sometimes evident
in dwarf starbursts \citep{schwartz04}.  The velocity of this gas
correlates weakly with star formation rate (SFR) and
galactic rotation speed, with a factor of $\sim30$ change in velocity
observed over a range of 4 orders of magnitude in SFR
\citep[$v_{\rm wind}\propto {\rm SFR}^{0.3}$;][]{rupke05b,martin05}. 

At intermediate redshift, the Mg II $\lambda\lambda$2796, 2803 ISM
absorption line shifts into the observed-frame optical.  The
relatively wide separation of this doublet makes it a good choice for
outflow studies. \citet{tremonti07} detected $500-2000$ ${\rm
  km~s}^{-1}$ outflow velocities in a small sample of $z \sim 0.5$
post-starburst galaxies, and suggested that these outflows cold be the
relics of AGN-driven winds.  \citet{weiner09} employed the spectral
stacking technique to probe $z \sim 1.4$ star forming galaxies drawn
from the DEEP2 survey.  This study, which is based on $\sim1400$
galaxy spectra, demonstrated that blue-shifted Mg II absorption is
ubiquitous in actively star forming galaxies.  

At high redshift, the observed-frame optical samples the rest-frame
far ultraviolet, which is rich in strong ISM resonance absorption
transitions, but lacking in stellar spectral features.  Because of the
difficulty of measuring the relative velocity of the stars and gas,
outflows have only been detected in very luminous Lyman break galaxies
\citep{pettini00, pettini02} or in composite spectra
\citep{shapley03}.

While much of our knowledge about galactic winds comes from studies of
local star forming galaxies, the work to date has been based on
relatively small samples of extreme objects (dwarf starbursts,
ULIRGs).  The properties of outflows in local normal star forming
galaxies are still largely unknown. In this paper, we investigate Na D
absorption in a sample of $\sim$150,000 star forming galaxies
drawn from the Sloan Digital Sky Survey. By stacking the spectra of
galaxies selected to have similar physical attributes, we obtain very
high S/N composite spectra. After carefully modeling and dividing out
the stellar continuum, we are able to probe very low Na D Equivalent
Widths (EWs) and extend our analysis over a wide range in galaxy
physical parameters.

  
This paper is arranged as follows. In \S2, we introduce the sample
selection criteria used in our study. Our method of creating composite
spectra and measuring the Na D lines is developed in \S3. We apply the
method to the SDSS sample in \S4 and summarize our results in \S5. We
use the cosmological parameters $H_0=70~{\rm km~s^{-1}~Mpc^{-1}}$,
$\Omega_{\rm M}=0.3$ and $\Omega_{\Lambda}=0.7$ throughout this paper.

\section{Sample selection}


\subsection{The Data \label{data}}

The Sloan Digital Sky Survey \citep[SDSS;][]{york00} spectroscopic
galaxy catalogue contains $\sim$930,000 spectra in its seventh data
release \citep[DR7;][]{abazajian09}. We analyze objects drawn from the 
``Main" galaxy sample \citep{strauss02}, which have Petrosian $r$ magnitude 
in the range $14.5 < r< 17.7$ after correction for foreground galactic 
extinction using the reddening maps of \citet{schlegel98}.
The spectra are taken through $3^{\prime\prime}$ diameter 
fibers and cover a wavelength range from 3800 to 9100\AA\ with a resolution of 
$R\sim2000$ and a dispersion of 69~km~s$^{-1}$~pixel$^{-1}$. Most of 
the galaxies have redshifts in the range $z=0-0.3$. We adopt redshifts 
from the \texttt{specBS} pipeline \citep{adelman08}. The median 
redshift error is $\sim 10^{-4}$. 

As described in \citet{tremonti04} and \citet{brinchmann04}, a stellar
population model is fit to the continuum of each galaxy spectrum after
masking out the strong nebular emission lines.  The basic assumption
is that any galaxy star formation history can be approximated as a sum
of discrete bursts. The library of template spectra is composed of
single stellar population (SSP) models generated using a preliminary version
of the population synthesis code of Charlot \& Bruzual (2010, in prep.)
which incorporates the MILES empirical stellar library
\citep{sanchez06}. We refer to these models hereafter as the `CB08
models'.  The spectral resolution (2.3~\AA\ FWHM), spectral-type
coverage, flux-calibration accuracy and number of stars in the MILES
library represent a substantial improvement over previous libraries
used in population synthesis models.  We generate template spectra
with ten different ages (0.005, 0.025, 0.1, 0.2, 0.6, 0.9, 1.4, 2.5,
5, 10 Gyrs) and four metallicities (0.004, 0.008, 0.017, 0.04). For
each metallicity, the ten template spectra are convolved to the
measured stellar velocity dispersion of each individual SDSS galaxy,
and the best fitting model spectrum is constructed from a non-negative
linear combination of the template spectra, with dust attenuation
modeled as an additional free parameter. The metallicity which yields
the minimum $\chi^2$ is selected as the final best fit. The results of
this fitting procedure and measurements of an number of line indices
(e.g., Lick\_Mgb, D4000) are made available in the MPA/JHU
catalogue\footnote{The MPA/JHU catalog can be downloaded from
  http://www.mpa-garching.mpg.de/SDSS/DR7.}.  The catalog also
includes the equivalent width (EW) and flux of an number of 
emission and ISM absorption lines (e.g., H$\alpha$, H$\beta$, Na D).

The SDSS imaging data consist of CCD imaging in $u, g, r, i, z$ bands
\citep{fukugita96, smith02}, taken with a drift scan camera
\citep{gunn98} mounted on a wide-field 2.5-m telescope. The SDSS
  photometric pipeline \citep{lupton01} fits each galaxy image with a
two-dimensional model of a \citet{devau48} surface profile and an
exponential profile, each convolved with the PSF of the image.  A
seeing-independent axial ratio ($b/a$) is derived by this fitting
procedure.  The pipeline also computes the best linear combination of
the exponential and de Vaucouleurs models and stores it in a parameter
called $\rm fracDeV$ \citep{abazajian04}.  Here we follow
\citet{padilla08} in using $\rm fracDeV$ to distinguish disk galaxies
(${\rm fracDeV}<0.8$) from early-type galaxies (${\rm
  fracDeV}\ge0.8$).  The axis ratios from the exponential and de
Vaucouleurs models are consistent with each other independent of the
morphology of the galaxy.  In this work, we are only concerned with
disk galaxies, and therefore we adopt the axis ratios from the
exponential fit.  We compute the galaxy inclination, $i$, from the
measured axial ratio, $b/a$, and the $r-$band absolute magnitude
$M_{\rm r}$ using Table 8 in \citet{padilla08}.  This empirical
relation takes into account the increase in bulge-to-disk ratio with
absolute magnitude.  Typical inclination errors are  
$10^{\circ}$ at $i > 40^{\circ}$.  At $i < 40^{\circ}$ galaxy
asymmetries can result in over-estimates of the inclination.
The photometric properties used in this paper
($\rm fracDeV$, $b/a$, $u, g, r, i, z$ magnitudes) are also available
in catalogs on the MPA/JHU webpage.

The derived galaxy parameters required in this work include stellar
mass ($M_*$ for the whole galaxies and $M_{\rm fiber}$ for stellar mass within 
fiber), stellar mass surface density ($\Sigma_{M_*}$), $V-$band
dust attenuation ($\rm A_V$), SFR, SFR surface density ($\Sigma_{\rm
SFR}$), and the SFR per unit stellar mass or `specific SFR' (SSFR = SFR/$M_{\rm fiber}$).
The stellar mass-to-light ratios are obtained by comparing
$urgiz$ colors of galaxies to a large grid of CB08 model colors 
following the methodology described in \citet{salim07} to avoid aperture 
correction issues. This approach
differs from that of \citet{kauffmann03a} who used the spectral 
indices D$_n$(4000) and $\rm H\delta_A$ to constrain 
the mass-to-light ratios. However, the differences between the mass
estimates of the two methods are very small, typically 0.01 dex.

We calculate $\rm A_V$, the optical dust attenuation in magnitudes,
using the H$\alpha$/H$\beta$ ratio.  We assume an intrinsic
H$\alpha$/H$\beta$ ratio of 2.87 \citep{oster06}, a Milky Way
attenuation curve \citep{Cardelli89}, and $R=3.1$.  The nebular
attuenation is likely to overestimate the attenuation of the stellar
continuum \citep[c.f.,][]{Calzetti94}.

We derive the SFR from the dust extinction corrected H$\alpha$
luminosity using the formula of \citet{kennicutt98a}, dividing by a
factor of 1.5 to convert from a Salpeter to Kroupa initial mass
function \citep{Kroupa01}. $\Sigma_{\rm SFR}$ is defined as SFR/$\pi R^2$ and 
$\Sigma_{\rm M_*}=M_{\rm fiber}/\pi R^2$, where $R$ corresponds to the SDSS 
$1.5^{\prime\prime}$ aperture radius corrected for projection effects.

Except stellar mass, all of these derived parameters are calculated 
from quantities measured within the SDSS fiber aperture and we do not apply aperture
corrections.  We believe the locally determined physical parameters
will have the greatest influence on the local ISM kinematics traced by
Na D; star formation occurring in the outer disk of the galaxy is
unlikely to drive a circum-nuclear outflow. Another exception to our
policy of using locally derived physical properties is
Figure~\ref{fig_lit_compare} where we use global galaxy SFRs in order
to compare with results from the literature. The global SFRs are 
estimated by adding the SFR inferred from the emission lines 
measured in the fiber aperture \citep{brinchmann04} and the 
SFR estimated from the colors outside 
the fiber from fits to the 5-band photometry. The total SFRs are 
comparable the global SFRs in \citet{brinchmann04}, but for a small 
subset of galaxies an aperture correction problem identified 
by \citet{salim07} has been rectified.

\subsection{Sample Selection \label{sample_selection}}

The criteria used to select the parent sample used in our
analysis are the following:

\begin{enumerate}
\item redshift of $\rm 0.05 \le z \le 0.18$. With 
$3^{\prime\prime}$ diameter fibers, the SDSS spectra probe the central 
3--12 kpc in this redshift range. At the median redshift of $z=0.09$, 
the fiber probes the central 6~kpc. 

\item $r-$band $\rm fracDeV < 0.8$. This ensures that we are selecting 
disk galaxies, which makes the calculation of the inclination angle 
from $b/a$ possible. 

\item D$_n$(4000) $< 1.5$. This index measures the 4000~\AA\ break,
  and is a good indicator of the galaxy star formation history
  \citep{kauffmann03a}. Our cut selects galaxies with young stellar
  populations. It is just below the threshold identified by
  \citet[][D$_n$(4000)=1.55]{kauffmann03b} where a sharp
  change in galaxy structural properties occurs. 

   

\item $\log ($[OIII]/H$\beta) <  
0.61/\{\log ([$NII]/H$\alpha)-0.05\} +1.3$.
We exclude AGNs from our sample using the emission line ratio diagnostics
given in \citet{kauffmann03c}. 

\end{enumerate}

We refer to this sample hereafter as `Sample A'. It contains
140,625 galaxies. In \S4 we will justify the application of some
additional cuts to help isolate galaxies with measurable interstellar
Na~D. In later sections of the paper, we will refer to these
sub-samples as Samples B and C.

\begin{figure*}
\bc
\hspace{-1.6cm}
\resizebox{17cm}{!}{\includegraphics{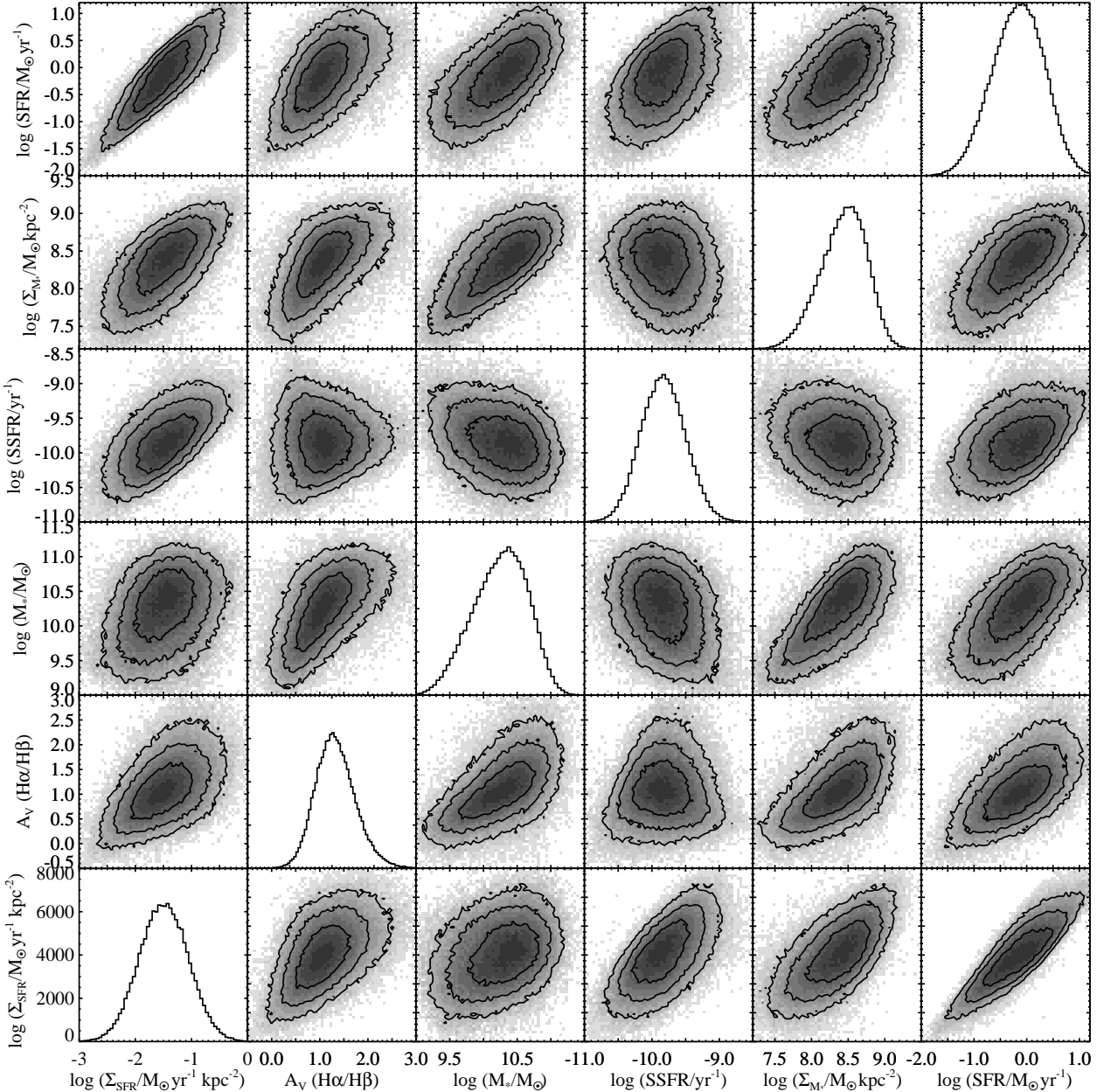}}\\%
\caption{Correlations among the physical parameters of our Sample
  A galaxies. For panels where the $x-$ and $y-$axis are the same, 
  histograms are shown. Note that all 
  quantities except $M_*$ are measured within the SDSS fiber aperture. 
  \label{Fig_param_correlations}}
\ec
\end{figure*}

In Figure~\ref{Fig_param_correlations} we illustrate the distribution of
galaxy physical properties in Sample A and show how these parameters
correlate with one another.  As we will discuss in
\S\ref{outflow_properties}, these correlations make it difficult to
distinguish the physical parameters that most strongly influence the
properties of galactic winds.
 
\subsection{Interstellar Na D in SDSS Spectra
\label{separating_NaD}}
Studying gas kinematics using optical absorption line spectroscopy is
challenging because of the lack of strong ISM absorption lines.  An
additional complication when using the galaxy continuum as the
background source, is the the prominence of stellar absorption
features in the continuum.  The Na~I~``D'' $\lambda\lambda5890,5896$
doublet is favored because it is moderately strong (sometimes
saturated) in starburst galaxy spectra \citep[c.f.,][]{heckman00}, and
because the stellar contamination is modest in galaxies dominated by
very young stars.  Other transitions which are sometimes used are
Ca~II~H \& K, which occur in a region of the spectrum rife with
stellar absorption and nebular emission lines, and 
K~I~$\lambda\lambda7665, 7699$ which is a relatively weak transition.

The main drawback of using Na~D to probe the ISM of normal star
forming galaxies is that it is a very prominent feature in the
the spectra of cool stars; stellar contamination can therefore be
significant. We discuss this issue further in \S\ref{continuum} where
we show that $\sim$80\% of the observed Na~D absorption in our
galaxy spectra arises in stellar atmospheres.

In most previous work, the stellar contribution to the Na I line was
estimated using the Mg~I $\lambda\lambda 5167, 5173, 5184$
triplet, which arises only in stellar atmospheres.  \citet{martin05}
find EW(Na~I) = 1/3 EW(Mg~I) based on an analysis of stellar spectra.
Here we take a somewhat more sophisticated approach, and model the
galaxy continuum light using stellar population synthesis models
(\S\ref{continuum}).  We use the Mg~I lines residuals to gauge the
success of our continuum fitting procedures.

\begin{figure}
\bc
\hspace{-0.6cm}
\resizebox{8.5cm}{!}{\includegraphics{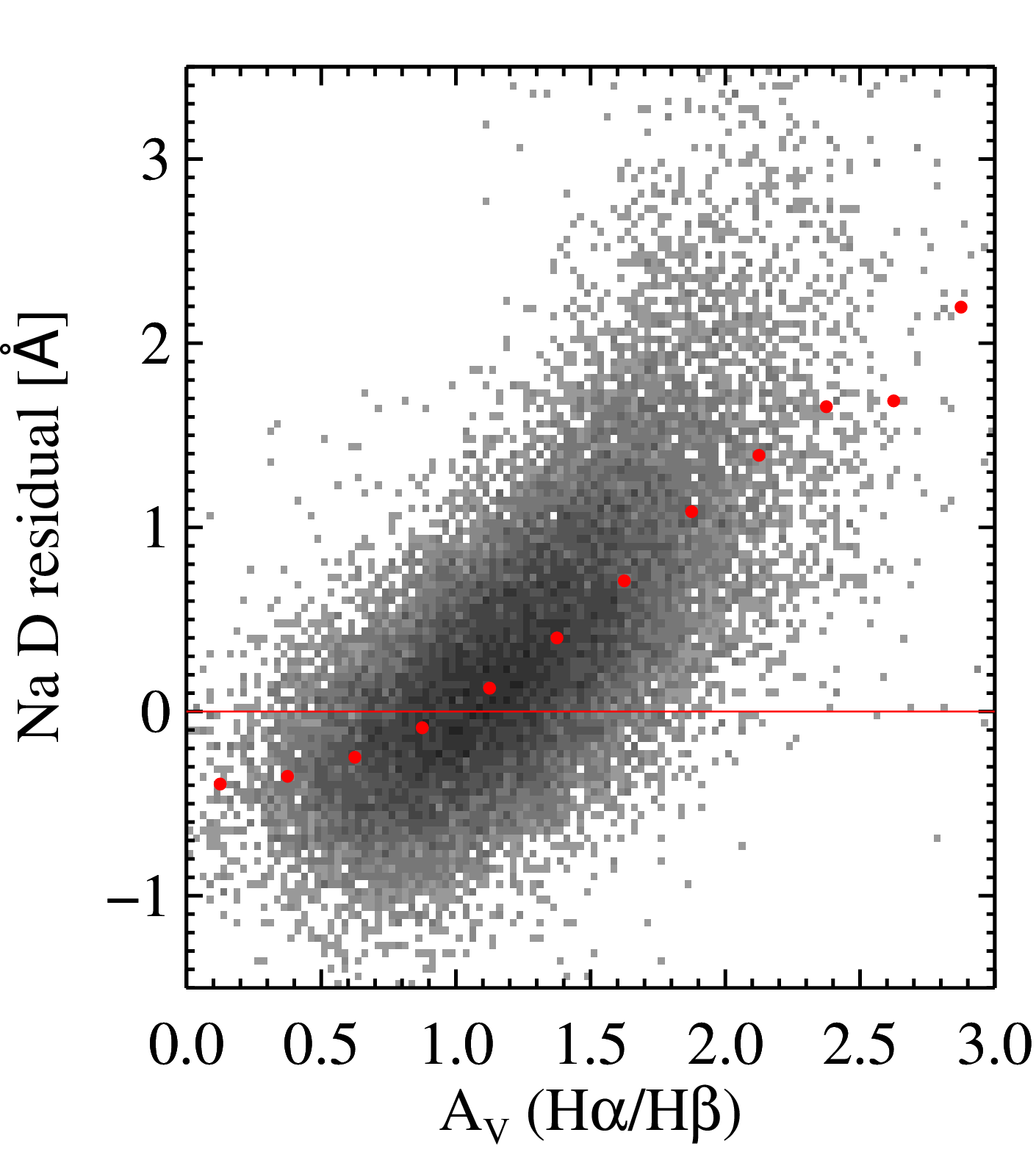}\includegraphics{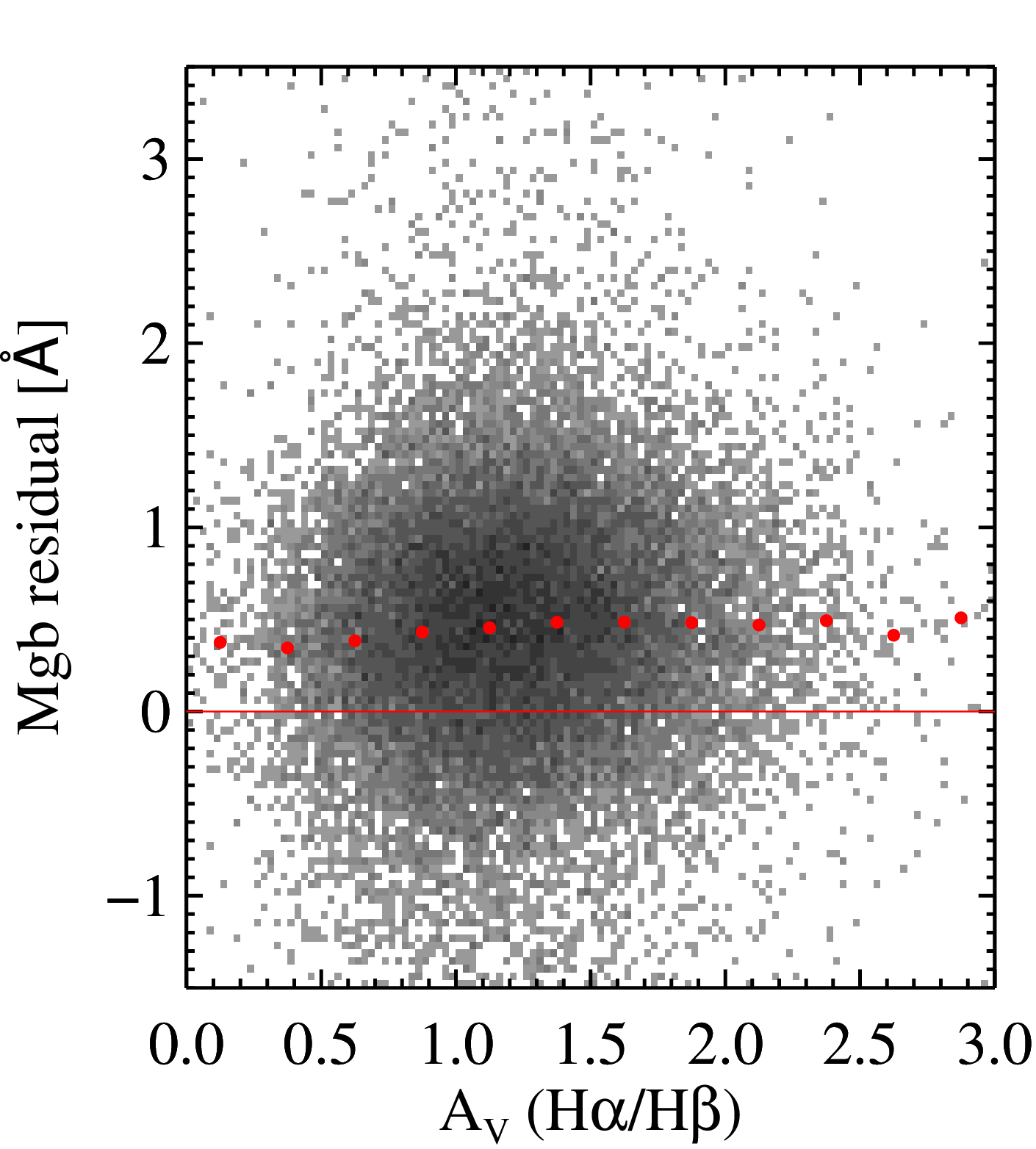}}\\%
\caption{Residual Na D (left) and Mg $b$ (right) absorption versus
  dust attenuation for Sample A galaxies with S/N $>$ 15. Red dots
  indicate the median. The residual absorption is measured after
  subtraction of the best fit stellar continuum model.  Dust
  attenuation is measured from the H$\alpha$ and H$\beta$ nebular
  lines. The strong positive correlation in the left panel is evidence
  that the excess Na D absorption arises in the dusty ISM.  The lack
  of correlation for Mg $b$ is expected since this is an excited
  photospheric feature that does not have a counterpart in the ISM.
\label{Fig_NaD_Av}}
\ec
\end{figure}

In our subsequent analysis we will make use of composite galaxy
spectra.  However, we begin by examining individual spectra in Sample
A with S/N pixel$^{-1} > 15$.  Roughly 19\% (27,076) of the
Sample A galaxies make this cut. For each galaxy we measure the Na D
EW and the Lick Mg~$b$ index \citep{worthey94} of the observed
spectrum and the best fit continuum model. In
Figure~\ref{Fig_NaD_Av} we plot the residual absorption ($data$ -
$model$) as a function of nebular attenuation.  We hypothesize that
the Na~D excess in the data relative to the models is due to ISM
absorption. Na~I has an ionization potential less than Hydrogen, so
dust shielding is expected to play an important role in its
survival \citep{murray07}.  The strong positive correlation between the Na~D residual
and dust attenuation is evidence that the excess Na D absorption
arises in the dusty ISM.  The lack of any corresponding correlation
in the Mg~$b$ residual indicates that the starlight is well fit by
our models even in heavily dust attenuated galaxies.  (The small
positive offset of the median is due to the 
presence of a weak [N~I] $\lambda$5198 emission in one of the continuum 
bands of the Mg~I index -- see Fig.~\ref{NaD_cont_example}).

One notable feature of Figure~\ref{Fig_NaD_Av} is tendency for
galaxies with low dust attenuation to have negative Na~D residuals --
more Na~D absorption in the stellar population models than in the
data.  There are three possible explanations for this: (1) the
S/N of the spectra are not high enough for a robust continuum fit;
(2) the stellar population models are not robust around the Na D
region, at least for some stellar populations; (3)  Na~D is 
primarily in emission in gas and dust-poor galaxies. Through
the spectral stacking technique, the S/N issue can be easily
resolved.  We discuss the other possibilities further in 
\S\ref{measuring_NaD}.


\subsection{Galaxies with Strong ISM Na D Absorption \label{frac_NaD}}

In the previous section we found that the Na D residual increases with
$\rm A_V$. In this section, we study how the Na D residual evolves
with the other galaxy parameters that we are interested in: SFR,
$\Sigma_{\rm SFR}$, $M_*$, $\Sigma_{M_*}$, SSFR.  We define galaxies
with a Na D residual $> 0.8$~\AA\ as strong ISM Na D absorption
galaxies. This definition is somewhat arbitrary, however, our
conclusions are not very sensitive to the threshold we adopt.

\begin{figure}
\bc
\hspace{-0.1cm}
\resizebox{8.5cm}{!}{\includegraphics{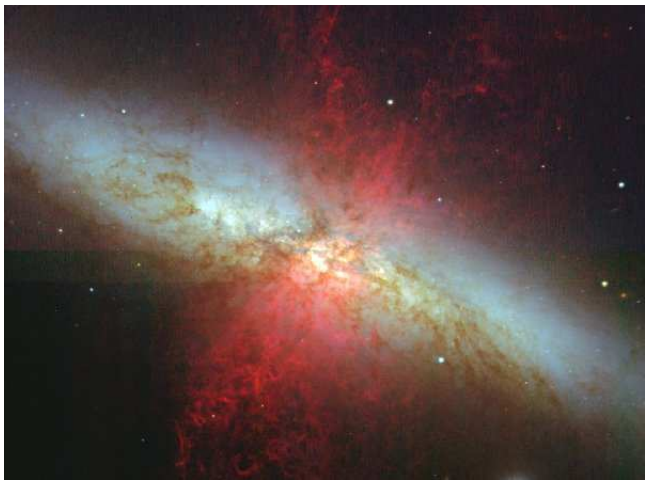}}\\%
\caption{The galactic wind in the prototypical starburst galaxy
    M82.  The color-composite image was constructed by the Hubble
    Heritage team from B,V,I, and H$\alpha$ images
    obtained with the \emph{Hubble Space Telescope} Advanced
    Camera for Surveys \citep{mutchler07}.  The galaxy has an 
    inclination of $i \approx
    80\hbox{$^\circ$ }$.  Outflowing gas is visible as a bi-polar
    cone of H$\alpha$ emission (red) extending along the galaxy minor
    axis.  If this wind geometry is common in disk galaxies, then we
    would expect to observe blueshifted Na~D when our sightline probes
    the wind bi-cone at $i < 60^{\circ}$.  The dusty filaments evident
    in the disk may give rise to Na D absorption at the galaxy
    systemic velocity.
    \label{Fig_M82}}
\ec
\end{figure}

Deep images of the edge-on starburst galaxy M82 in the X-ray,
H$\alpha$, and 8$\mu$m-bands \citep{ohyama02, mutchler07, engel06}
indicate that gas and dust are being driven out along the galaxy minor
axis in a weakly-collimated bipolar outflow (See
Figure~\ref{Fig_M82}). Imaging of the H$\alpha$ and soft X-ray emission 
shows that this wind geometry is common in starburst galaxies 
\citep[e.g.,][]{lehnert96, strickland04}.
Thus, absorption line studies of nearly
face-on systems (low~$i$) should probe gas in the outflow, while
observations of nearly edge-on systems (high~$i$) will not; they may,
however, be sensitive to gas in the disk.  In keeping with this
picture, \citet{heckman00} found that there is a high probability
($\sim$ 70\%) of detecting outflowing gas in absorption in starburst
galaxies with inclinations less than $60\,^{\circ}$. Based on these
results, we split our Sample A into two sub-samples with $i <
60\,^{\circ}$ and $i > 60\,^{\circ}$.

\begin{figure}
\bc
\hspace{-0.6cm}
\resizebox{8.5cm}{!}{\includegraphics{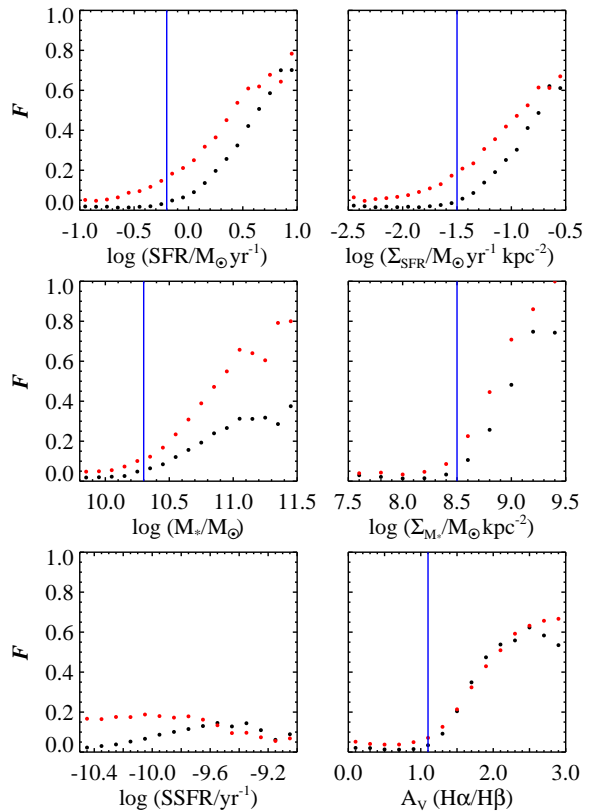}}\\%
\caption{The fraction of strong Na D residual galaxies as a function
  of different galaxy parameters.  The fraction is measured as $F = N({\rm
  EW}>0.8)/N_{\rm bin}$, where $N({\rm EW}>0.8)$ is the number of
  galaxies with Na D residual $> 0.8$~\AA\ in a given bin and $N_{\rm
  bin}$ is the total number of galaxies in the bin.  
  The black dots represent galaxies in Sample A with $i <
  60\,^{\circ}$ and red dots represent the 
  galaxies with $i > 60\,^{\circ}$. The blue vertical line in each panel marks 
  the place where the fraction of $i < 60\,^{\circ}$ galaxies with
  strong Na~D residual reaches 5\% and begins to rise steeply. 
\label{Fig_frac_NaD}}
\ec
\end{figure}

Figure \ref{Fig_frac_NaD} shows how the fraction of galaxies with Na~D
residuals greater than 0.8~\AA\ changes with various galaxy parameters
for our two samples split by inclination.  The fraction increases
rapidly with all of the galaxy properties, except SSFR, for both low
(black) and high (red) inclination samples. The highly inclined
galaxies have stronger Na D residuals; in \S\ref{inc_effect} we will
argue that this strong Na D absorption arises in the disk.
Galaxies with strong ISM Na D absorption make up only a small part of
the full sample; however, they dominate the populations with strong
star formation, high stellar mass, and high dust attenuation. The blue
vertical line in each panel marks the place where the fraction of
galaxies with $i<60\,^{\circ}$ and strong Na D reaches 
5\% and begins to rise steeply.

\section{Data analysis \label{data_analysis}} 

In this section, we describe how we measure the properties of the ISM
Na D absorption. As discussed in \S\ref{separating_NaD}, the low S/N
of the individual spectra will lead to large uncertainties in the
measured Na~D EW.  The median Na D EW error for Sample~A is
$\sim$1.5~\AA, which is greater than the strength of the interstellar
absorption in most galaxies.  By stacking (averaging) galaxy spectra
we avoid this S/N issue and are able to use samples that are
relatively complete.  The steps of our analysis are as follows:
 
\begin{itemize}
\item Stack spectra in different bins of galaxy physical
parameters to create high S/N composite spectra.  The parameters used
in our stacking analysis are galaxy inclination ($i$), $M_*$, 
$\Sigma_{M_*}$, SFR, $\Sigma_{\rm SFR}$, SSFR, and $\rm A_V$. 
\item Fit the CB08 stellar population models to the stacked spectra 
generated in the first step. The aim is to separate stellar and ISM 
Na D absorption.  Divide each spectrum by the best fitting 
continuum model.
\item Fit the interstellar Na D line in each continuum-normalized
spectrum. Return the EW, line centroid, 
line width, optical depth, and covering factor.
\item Look for correlations between the properties of the gas and the
galaxy physical properties.

\end{itemize}
Each of these steps is described in more detail below.

\subsection{Stacking galaxy spectra\label{stacking}}

The galaxy spectra are first corrected for foreground Galactic
attenuation using the dust maps of \citet{schlegel98}, transformed
from vacuum wavelengths to air, and shifted to the restframe using the
redshift determined by the SDSS \texttt{specBS} pipeline. We then
normalize each spectrum to its median flux between 5450 to 5550\AA,
where the spectrum is free of strong absorption and emission lines.  The
normalized restframe spectra in a given parameter bin are averaged 
(``stacked'') using the following weighting scheme:
\begin{equation}
f_{\rm comp}(\lambda)=\frac{\sum_{i=1}^n {mask(i,\lambda) \times  f(i,\lambda)}}{\sum_{i=1}^n {mask(i,\lambda)}}
\end{equation}
where the sum is over $i=1,2,...,n$ galaxies with normalized flux
$f(i,\lambda)$. We set $mask(i,\lambda)=0$ for bad pixels (identified in the
SDSS mask array), and $mask(i,\lambda)=1$, otherwise. This method
gives equal weight to all spectra, but excludes bad pixels from the composite. 

We adopt an adaptive binning approach to produce composite spectra in
different parameter bins with equal S/N.  The galaxies are sorted by
the physical parameter of interest and added to the stack one at time, 
from lowest to highest.  After each addition, the S/N of the continuum
near Na~D is computed. The process is repeated until a S/N of 300 is
reached. The value of the physical parameter of 
interest in each bin is computed from the median of the individual
measurements.  

\subsection{Fitting the Continuum  \label{continuum}}

The aim of this study is to use the ISM Na D absorption to probe
galactic winds in star forming galaxies. However, as discussed in
\S\ref{separating_NaD}, we can not disregard the fact that
interstellar Na~D absorption is superimposed on stellar Na~D
absorption.  In previous studies of LIRGs and ULIRGs \citep{martin05,
  martin06, rupke05a, rupke05b}, stellar Na~D absorption was estimated
from Mg~I, and deemed to contribute only a small fraction ($<10$\%) to
the total line EW.  However, in the investigation of 32 far-IR-bright
starburst galaxies, \citet{heckman00} found that $>40$\% of the
galaxies (mostly edge-on systems) had strong stellar contamination.  
It is well-known that Na~D is strong in the spectra of cool
stars, with a peak strength in the range from K3 to M0
\citep{jacoby84}. These stellar types are expected to make a
significant contribution to the current sample since the SDSS spectra
are obtained through a $3^{\prime\prime}$ circular fiber aperture
that samples the central 3 -- 12 kpc of our current sample.  In these
regions, bulge K-giants are likely to make a strong contribution to the
integrated the stellar light \citep{heckman80, bica91}.

To separate the stellar and ISM contribution to the Na~D absorption we
model the stellar continuum of each stacked spectrum using the CB08
stellar population synthesis models. In the Appendix, we show
that our results are insensitive to the choice of stellar population
models. Our continuum fitting procedure is nearly the same as that
used in fitting individual spectra (see \S\ref{data}). The main
difference between the individual and stacked spectra is that the
stacked spectra have much higher S/N. This makes it possible for us to
constrain the stellar contribution using a larger number of SSP
templates. We therefore fit the stacked spectra with all 40 templates
(4 metallicities and 10 ages) simultaneously. This allows the best-fit
model to consist of a mixture of metallicities, unlike the situation
for the individual spectra (\S\ref{data}).

The He~I $\lambda$5876 nebular emission line on the blue side of Na~I
$\lambda$5890 is broad enough to slightly influence our measurement of
Na~D.  Since the line presumably originates in H~II regions behind the
Na~D absorbing gas, we include it in our continuum model.  We fit a
Gaussian to the blue wing of He~I emission and add a model of the full
line to the stellar continuum before normalizing the spectrum.

\begin{figure*}
\bc
\hspace{-1.6cm}
\resizebox{17cm}{!}{\includegraphics{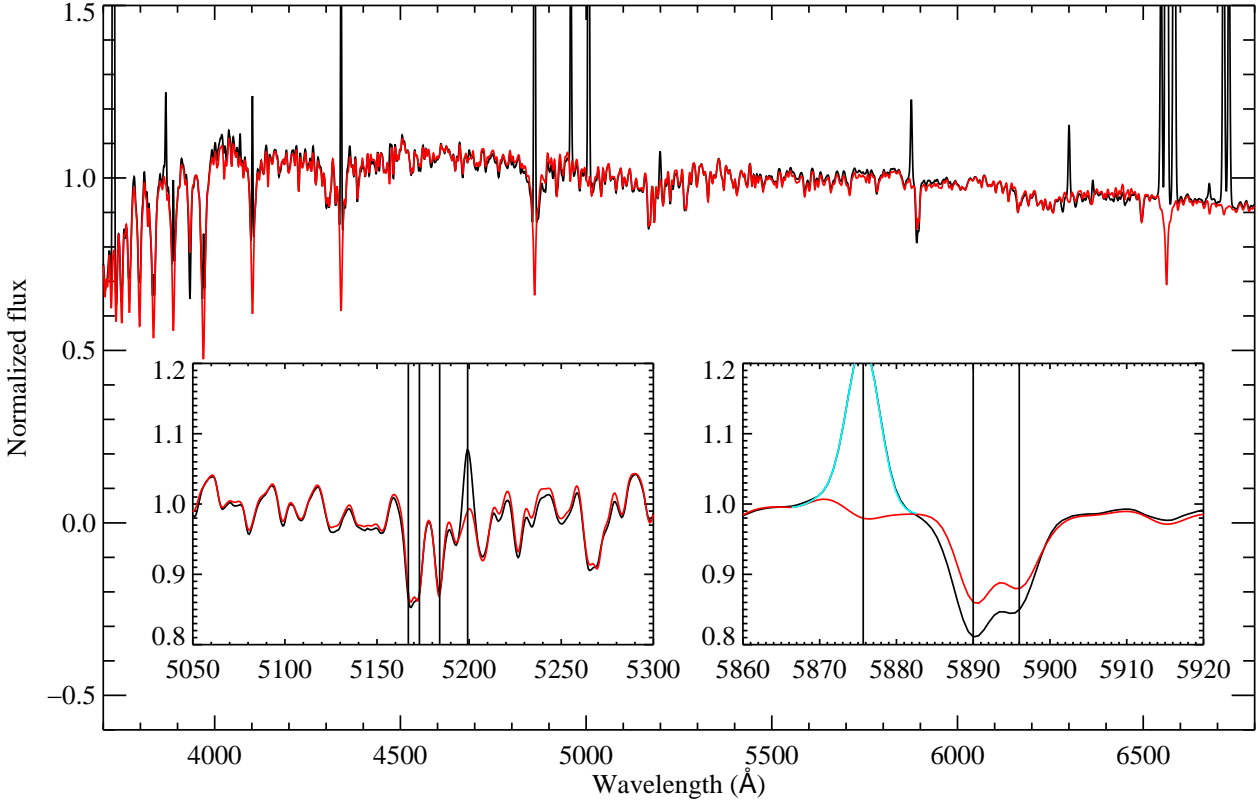}}\\%
\caption{An example of a stacked spectrum (black) and its best-fit 
continuum model (red).  The spectrum is a composite of 2000 randomly
selected galaxies in Sample~A with 
$\Sigma_{\rm SFR} > 0.1 M_\odot~{\rm yr}^{-1}~{\rm kpc}^{-2}$. 
The two insert drawings highlight regions near the Mg~I (left) and 
Na~I lines (right). The 
vertical black lines mark the restframe wavelength of the Mg~I
$\lambda\lambda$5167, 5173, 5184 triplet, the [N~I] $\lambda$5198, 
the He I $\lambda$5876 emission line, and the 
Na~I $\lambda\lambda$5890, 5896 doublet. A Gaussian fit to the He~I emission 
line is shown in cyan.
\label{NaD_cont_example}}
\ec
\end{figure*}

Observationally, in local galaxies, galactic winds are a general
consequence of high SFR or $\Sigma_{\rm SFR}$ (Lehnert \& Heckman
1996); they are common in galaxies above a $\Sigma_{\rm SFR}$
threshold of 0.1 $M_\odot~{\rm yr}^{-1}~{\rm kpc}^{-1}$ (Heckman
2002).  To illustrate our continuum and absorption line fitting
procedure, we randomly select 2000 galaxies with $\Sigma_{\rm SFR} >
0.1 M_\odot~{\rm yr}^{-1}~{\rm kpc}^{-2}$ from Sample A and stack
them. Figure~\ref{NaD_cont_example} shows the resulting stacked
spectrum (black line) and our stellar continuum fit (red line). We
highlight the regions near Mg I (left insert drawing) and Na~I (right
insert).  The vertical black lines mark the restframe wavelength
center of the Mg I $\lambda\lambda$5167, 5173, 5184 triplet, [N~I] $\lambda$5198, 
He I $\lambda$5876, and the Na I $\lambda\lambda$5890, 5896 doublet. A
Gaussian fit to the He I emission line is shown in cyan.  From this
plot, we infer that: (1) the stellar continuum model provides a very
good fit overall; (2) the stellar contribution of the Na~D absorption
is as high as $\sim$80\% on average.  Since the full Sample A includes
galaxies with lower $\Sigma_{\rm SFR}$ and a reduced incidence of
strong ISM Na~D absorption, we expect the stellar contribution to the
Na~D line to be even larger in some cases.

\subsection{Measuring Interstellar Na~D Absorption Line Profiles 
\label{measuring_NaD}}

There are two common methods of measuring absorption line profile
parameters: (1) fitting simple functional forms (such as Gaussian or
Voigt profiles) to the intensity as a function of wavelength; (2)
making more complex models, in which the intensity profiles are direct
functions of physical parameters \citep[velocity, optical depth, and
covering fraction; e.g.,][]{rupke05a}. The first solution is mainly
based on mathematics rather than physics, and it is widely used for
single lines or unblended doublets or multiplets.
The study of unblended transitions is much easier than blended ones.  For
the blended doublet or multiplet, we prefer the second method since on
one hand, the profile shape is readily understood in terms of a set of
physical parameters, and on the other hand, different model assumptions
can be tried.  For the blended Na doublet, we choose to use the second
method. We briefly summarize Rupke's model below and refer the reader
to \citet{rupke05a} for more details.

Previous studies of Na~D absorption troughs in ULIRGs strongly suggest
the assumption of complete continuum coverage is not met
\citep{martin05, martin06, rupke05a, rupke05b}.  These studies also
found super-thermal Na~D line widths ($b\sim300$~km~s$^{-1}$), which
can arise from multiple fragmented shells of cool gas traveling at
different velocities \citep{fujita09}.  When stacking many hundreds of
galaxy spectra, absorption lines with different velocities, line
widths, optical depths and covering factors will be blended.
\citet{Jenkins86} explored this situation 
extensively using monte-carlo simulations, and determined that
reasonable results can be obtained for the ensemble properties of the
absorbers provided the distribution function of their individual
properties is not too unusual. Results from stacked spectra may, in
fact, be more robust than those from individual galaxies, because the 
number of blended components is larger.  

\begin{figure*}
\bc
\hspace{-1.6cm}
\resizebox{17cm}{!}{\includegraphics{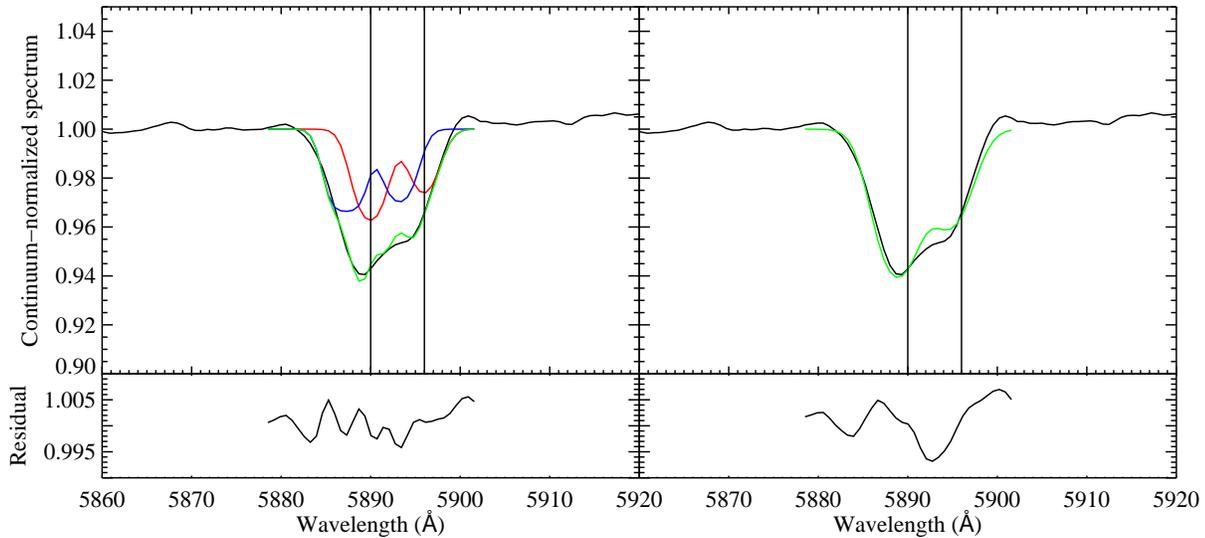}}\\%
\caption{An example of our Na~D line profile fits.  The back line in
  the top panel shows the spectrum from Figure \ref{NaD_cont_example}
  normalized by its best fit continuum model.  The left panel
  illustrates our two-component Na~D fit: the red line represents the
  systemic component; the blue line is the outflow (blue-shifted)
  component; and the green line shows the combined fit. For
  comparison, we also show a single component fit in the right
  panel. The residuals of are shown in the bottom panels.  The two
  component fit does a better job of modeling the shape of the line
  profile.  }
\label{fit2nad_exam}
\ec
\end{figure*}

We adopt the partial covering model for the current study and assume a
covering factor, $\cf$, which is independent of velocity.  In this case,
the general expression for the intensity of a doublet, $I(\lambda)$,
assuming a continuum level of unity, is
\begin{equation} \label{case1}
I(\lambda) = 1 - \cf + \cf e^{-\tau_B(\lambda)-\tau_R(\lambda)}.
\end{equation}
where $\tau_B(\lambda)$ and $\tau_R(\lambda)$ are the optical depth of
the blue and red components of the doublet, respectively. For Na~D,
$\tau_B/ \tau_R$=2.  Under the curve-of-growth assumption, the optical
depth $\tau$ of a line can be expressed as:
\begin{equation} \label{cog}
\tau(\lambda) = \tau_0 e^{-(\lambda-\lambda_0)^2/(\lambda_0 b/c)^2},
\end{equation}
where $\tau_0$ and $\lambda_0$ are the central optical depth and 
central wavelength of the line and $b$ is the line width.
The four physical parameters of this model are: 
\begin{itemize}
\item Outflow velocity ($v_{\rm off}$), which measures the
shift of line profile centroid relative to the galaxy's 
systemic velocity ($v_{\rm sys}$). 

\item Optical depth at line center ($\tau_0$) of the absorbing clouds, 
which influences the depth and intensity ratio of the two lines of the doublet.

\item Covering factor ($C_f$) of the absorbing clouds, which 
determines the residual flux at line center.

\item The Doppler width ($b$) of the absorption lines. 
\end{itemize}
Given the low resolution of our spectra ($\sim150$~km~s$^{-1}$ FWHM) 
and the blended nature of the profiles, there are some degeneracies 
between optical depth and covering factor which both influence the 
depth of the lines. The line width and the outflow velocity are more 
robust to degeneracies.

In the top panel of Figure~\ref{fit2nad_exam} we show the stacked
spectra of Figure~\ref{NaD_cont_example} after continuum
normalization.  The rest wavelengths of the two sodium lines are
marked by vertical lines.  The line profiles appear asymmetric,
with some of the gas blue-shifted relative to the systemic velocity of
the galaxy. We hypothesize that there is an absorption component at
the systemic velocity that arises in the disk, and an outflow
(blue-shifted) absorption component that arises in a galactic wind.
Systemic and outflow absorption line components were identified in
previous studies of individual starbursts \citep[c.f.,][]{martin05},
and in composite spectra of intermediate redshift galaxies
\citep{weiner09}.  We consider the origin of the two line components
further in \S\ref{inc_effect}, where we examine the inclination
dependence of the Na~D line profile.
 
We fit the Na~D line profile with two absorption components, as shown in
Figure~\ref{fit2nad_exam}, with the velocity of one component fixed 
at $v_{\rm sys}$ (red). The blue line in Figure~\ref{fit2nad_exam} denotes the
outflow component. We also show the single component fit in the top-right 
panel (green). Comparing the residuals of these two fits, we find that the 
two component fit is superior.  To aid in breaking the degeneracy of the two 
components, we have also fixed the line width of the systemic component to 
the width of the He~I emission line. We assume that the He~I line profile reflects the
kinematics of H~II regions in the disk, and that the gas producing the
systemic  Na~D absorption shares these kinematics. To have an idea of
how the final results depend on this assumption, we tried two other
methods: (1) fixing the systemic component to 1.5 times the width of
the He~I emission line; and (2) letting it vary in the range of
[1--1.5] He~I width. While method (2) gives very similar fitting
parameters as that of fixing the width to He~I, method (1) produces
larger spectral residuals and the fitted parameters show more scatter
with galaxy physical properties.  We therefore fix the linewidth of
the systemic component to He~I.

As noted in \S\ref{separating_NaD}, there are cases where the best-fit
continuum model has stronger Na~D absorption than the actual data.
The reasons for this are still not fully understood.  One possibility
is that the young stars used in the empirical stellar libraries of
CB08 have some excess absorption at Na~D due to the Milky Way's ISM.
However, a comparison with fully theoretical models (see Appendix)
suggests that such absorption, if present, is not very strong. Another
possibility is that a combination of 40 instantaneous burst models
with a range in age and metallicity is insufficient to represent the
the true galaxy star formation history.  Degeneracies may also limit
the ability of our fitting code to arrive at the optimal solution.
  We cannot rule out continuum fitting errors, however, we find
  this explanation less likely, given the high quality of the fits
  elsewhere in the spectrum (c.f., Fig.~\ref{NaD_cont_example}).  
  The final possibility is that Na~D is
  sometimes in emission.  This is plausible because Na D is a
  resonance absorption transition. (Each absorbed photon is re-emitted
  isotropically.)  Whether Na~D is detected in absorption or
  emission in a given spectrum depends on the relative geometry 
  and velocity of the gas and stars probed by the fiber. We note that 
  the emission appears slightly redshifted and it is seen 
  predominantly in galaxies that are 
  face-on and have low dust attenuation. This suggests that we are
  seeing through the disk to the back side of the expanding bi-polar
  outflow.  As the disk becomes progressively more dusty or inclined
  the emission from the far side of the disk is attenuated. Emission
  was not detected in previous studies of LIRGs and ULIRGs due to
  the dusty nature of these sources.  However, \citet{phillips93} 
  saw a clear Na~D P-Cygni profile (blushifted absorption, redshifted
  emission) in the disk galaxy NGC 1808. In the DEEP2 Mg~II study, 
  \citet{weiner09} found Mg~II emission in a subset of the galaxies. 
  The emission tends to be in the bluest of the galaxies. In some of 
  them it is visible in the individual spectra, not just the stack. They 
  don't have inclination information for the high$-z$ galaxies, but it 
  is natural to think that bluer galaxies are lower inclination on average. 
  
We include the objects with emission Na~D residuals in the stacking
and accommodate the possibility of Na~D
emission by allowing the systematic component in our line profile fit to
be in either absorption or emission.  We note that in some cases a
substantial systemic emission component is fit, even when the line
profile appears to be completely in absorption (c.f., the first panel
of Figure~\ref{NaD_inc_fit}).  In such cases, large residuals result from imposing the
requirement that the systemic component be in absorption (note the
strong asymmetry of the line profile).   We believe these fits are
robust and represent real cases of superimposed Na~D emission from the
disk and absorption from the outflow.  Figure~\ref{Fig_NaD_inc_param} shows very smooth
trends with galaxy physical properties as the systemic component
transitions from absorption to emission, suggesting that our code
generally does a good job of distinguishing the outflow and systemic
components whether the latter is in absorption or emission.
We have therefore chosen to allow the systemic
component to be in emission; for simplicity, we implement this by
allowing for negative covering factors.

\section{Probing galactic winds using stacked spectra \label{probe_winds}}

The dependence of the outflow properties on galaxy physical properties
is of great interest because it can help clarify the influence of SFR
and stellar mass on the wind kinematics.  We can also learn which
galaxies host outflows and thereby gain insight into the
origin of metals in the IGM. These empirical results can, in turn, be
incorporated into numerical simulations of galaxy formation and
evolution.

The data analysis method described in \S3 is applied into our star
forming galaxy sample in this section. We explore the effect of
changes in galaxy inclination in \S\ref{inc_effect}, and examine how
the outflow and systemic components of the ISM Na~D absorption evolve
with various galaxy physical properties in
\S\ref{outflow_properties}. This is the first time outflow properties
have been studied with large complete sample in the local universe
that includes normal star forming galaxies as well as starbursts.

\subsection{Trends with Galaxy Inclination \label{inc_effect}}

It has been suggested that the outflow component of the Na~D
absorption arises from ambient interstellar material that has been
entrained and accelerated along the minor axis of the galaxy by a hot
starburst-driven superwind \citep{heckman00}, while the systemic
component of the absorption arises in the disk.  Based on this
simple picture, we expect the two Na~D components to have strong -- and
opposite -- variations with galaxy inclination. We test our
assumptions about the absorber geometry using spectra stacked in
bins of galaxy inclination. 

\begin{figure}
\bc
\hspace{-0.6cm}
\resizebox{8.5cm}{!}{\includegraphics{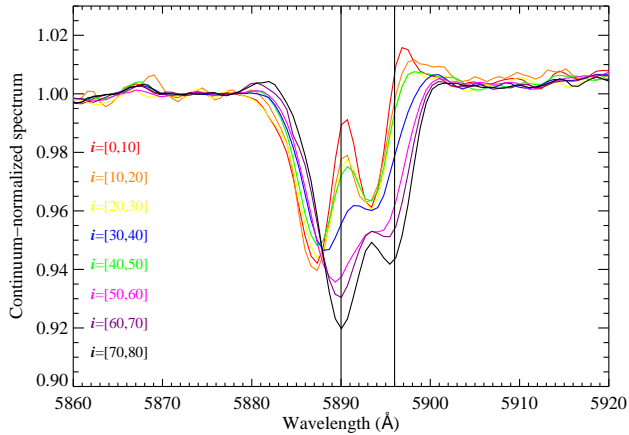}}\\%
\caption{Continuum-normalized spectra binned by galaxy inclination. 
Different inclination bins are indicated by different colors, with the
galaxies closest to face-on in red and the galaxies closest to edge-on
in black.
\label{NaD_res_inc}}
\ec
\end{figure}

Galactic winds are ubiquitous in galaxies with 
$\Sigma_{\rm SFR} > 0.1 M_\odot~{\rm yr}^{-1}~{\rm kpc}^{-1}$.
To test our picture of the wind outflow geometry, we 
therefore apply this $\Sigma_{\rm SFR}$ cut to our `Sample A' galaxies, to 
select sources likely to host outflows. 
In Figure~\ref{NaD_res_inc} we compare the interstellar Na D absorption 
profiles for these galaxies in 8 different inclination bins. Three
features are immediately obvious. 
First, the line profile depth at the rest wavelength of the doublet
increases dramatically as inclination increases and galaxies are
viewed more nearly edge-on.  Second, the line center shifts to bluer
wavelengths as inclination decreases and galaxies are viewed more nearly
face-on. Third, the lines are not saturated --- in which case the
doublet ratio would be closer to 1:1 -- nor are the profiles black at
the line center, which would indicate complete covering of the source
by the absorber.

\begin{figure*}
\bc
\hspace{-1.6cm}
\resizebox{17cm}{!}{\includegraphics{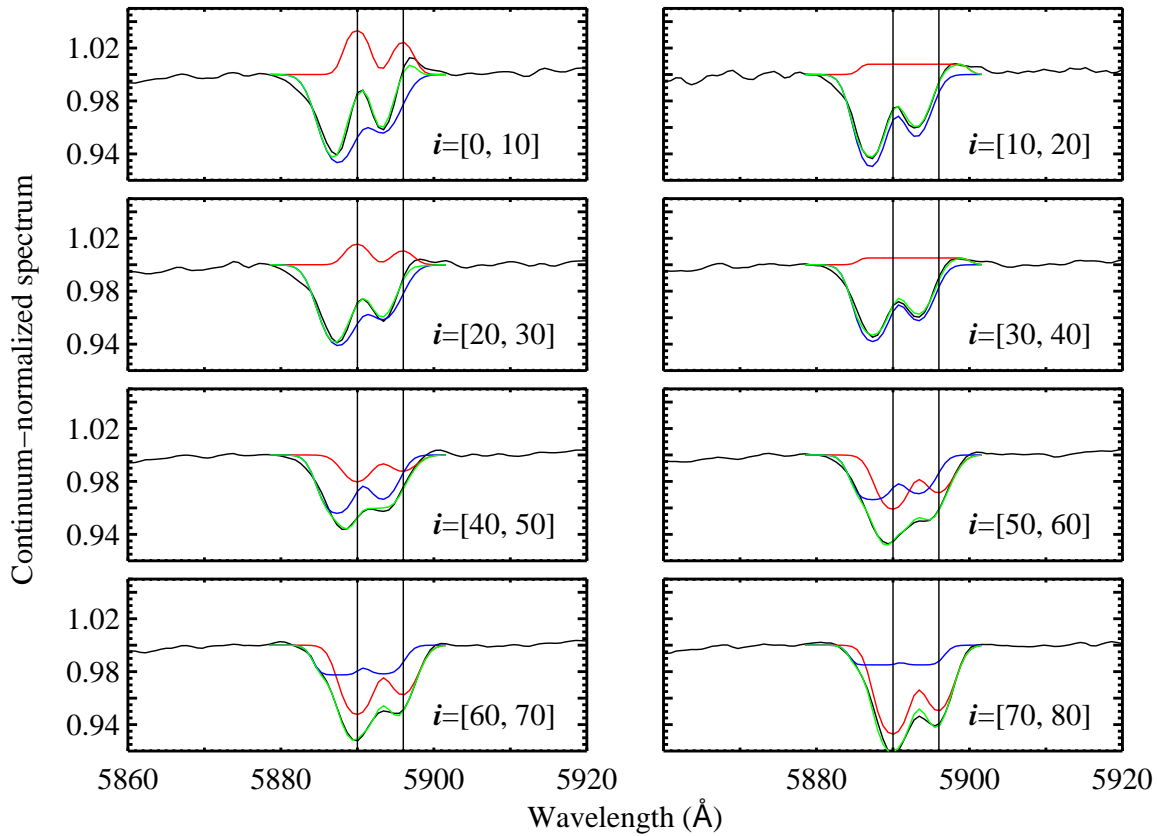}}\\%
\caption{Absorption line profile fits to composite spectra in 
different inclination bins. The black lines show the continuum
normalized composite spectra; the red line shows the absorption line
component fixed to the galaxy systemic velocity; the blue line
shows the outflow component; and the green line shows the
combined fit.  Note the appearance of Na~D emission at the systemic
velocity in the low inclination bins. 
\label{NaD_inc_fit}}
\ec
\end{figure*}

Figure \ref{NaD_inc_fit} illustrates our two-component absorption line
fits to composite spectra in different inclination bins. The red lines
show the component fixed at $v_{\rm sys}$, the blue lines show the
outflow component, and the green lines show the combined fit.  
The systemic component (red) evolves markedly from emission at low
inclinations to become the dominant absorption component at high
inclinations.  The outflow component (blue) evolves strongly in the
opposite sense, dominating the line profile in face-on systems, but
becoming a minor contributor to the absorption in edge-on galaxies.

\begin{figure*}
\bc
\hspace{-1.6cm}
\resizebox{17cm}{!}{\includegraphics{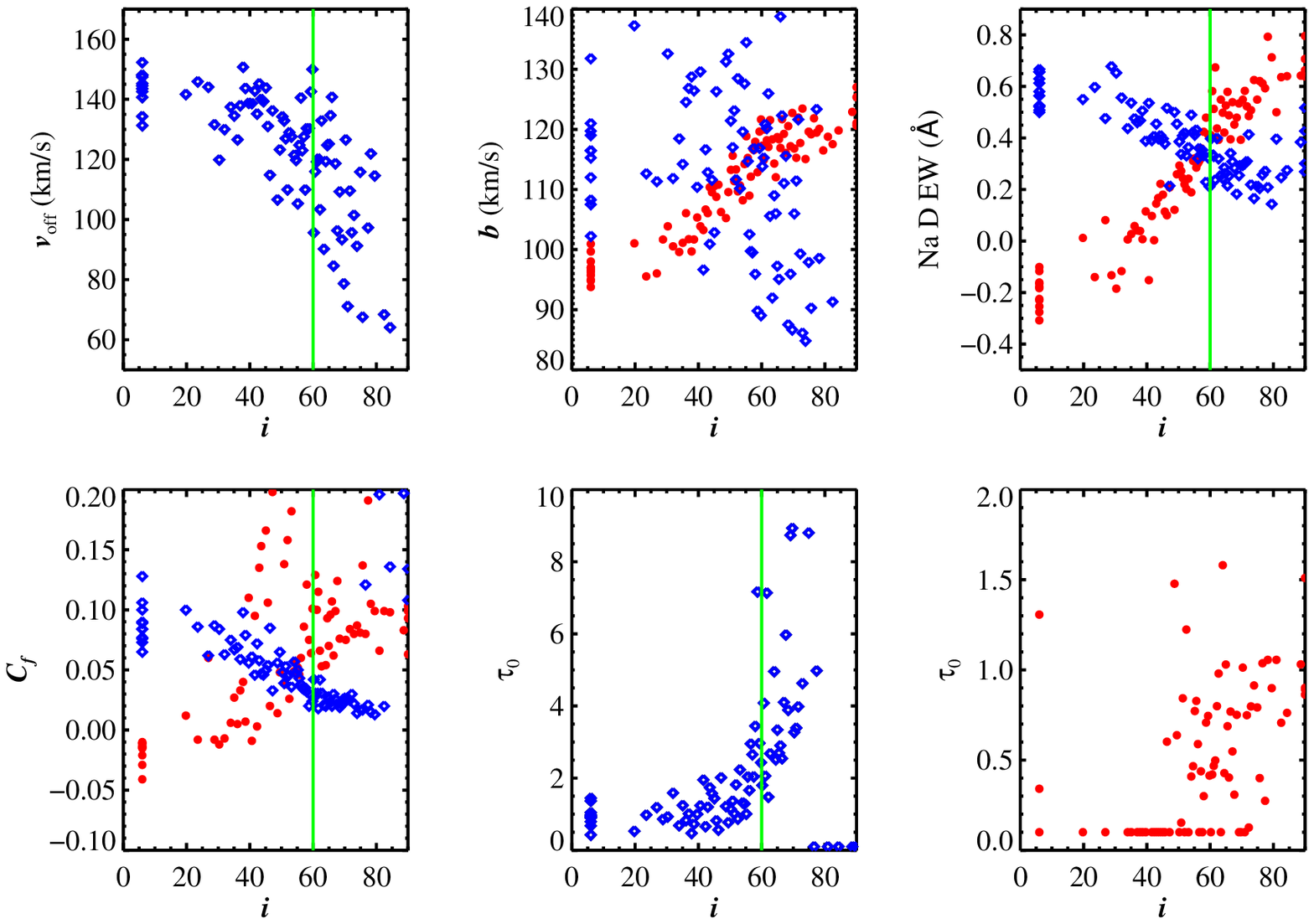}}\\%
\caption{Derived parameters of the outflow and systemic 
Na~D absorption components as a function of inclination angle.  
The blue diamonds represent the outflow component and the 
red circles represent the systemic component.  The parameters
shown include the velocity of the outflow component ($v_{\rm off}$), 
the line width ($b$), the covering factor ($C_f$), the optical 
depth at line center ($\tau_0$) and the line equivalent width (EW).
\label{Fig_NaD_inc_param}}
\ec
\end{figure*}

The equivalent width (EW), velocity ($v_{\rm off}$), line width ($b$),
covering factor ($C_f$) and optical depth at line center ($\tau_0$) of
the two absorption line components are shown as a function of
inclination angle in Figure~\ref{Fig_NaD_inc_param}.  At $i <
60\,^{\circ}$, the outflow velocity depends only weakly on
inclination, and has a median value of $\sim$140 $\rm km~s^{-1}$.
Above $i = 60^{\circ}$, the velocity drops precipitously.  The line
width of the outflow component decreases by $\sim40$\% as galaxies go
from face-on to edge-on, but shows considerable scatter.  The EW of
the outflow component changes smoothly and nearly linearly from
EW=0.7~\AA\ to at $i\sim0^{\circ}$ to nearly 0.1 at $i=90^{\circ}$.
This trend appears to be driven primarily by a decrease in the gas
covering factor. The optical depth of the outflow component increases
sharply above $i=60^{\circ}$. This may indicate that the
wide-angle outflow of dusty material has a more transparent and
fast-moving ``core" ($i < 60^{\circ}$) and a surrounding sheath of
optically-thick slower-moving entrained material ($i > 60^{\circ}$)
-- sort of a ``turbulent boundary layer".  However, the outflow
component makes a relatively small contribution to the total EW in
this regime, so some caution in interpreting the result is advised.

The systemic absorption component changes with inclination in the
opposite sense of the outflow component: the line width increases with
inclination by $\sim20$\% and the EW increases, from emission at
$i\sim0^{\circ}$ to 0.7~\AA\ of absorption at $i\sim90^{\circ}$.
The lack of zero-velocity ISM absorption component at low inclinations 
is consistent with the result of \citet{weiner09}, in which they found 
that in the bluest subset, which are likely to be galaxies viewed face-on, 
there was no component of absorption at systemic velocity.
Again, most of this change can be attributed to the covering factor;
the optical depth shows little trend with inclination, and
considerable scatter.  
The scatter in Figure~\ref{Fig_NaD_inc_param} represents a lower bound 
   to the uncertainities in the fits.

The trends shown in Figure~\ref{Fig_NaD_inc_param} are in agreement with our
basic hypothesis that Na~D absorption arises in both the galaxy disk and in
an outflow directed along the galaxy minor axis.  Notably, the
strength of the systemic component increases as galaxies are viewed
more nearly edge-on and a greater path length through the disk is probed.
This leads us to conclude that the systemic component arises in the warm
neutral medium of the disk.  These disk clouds could either be in the
central star forming region, in which case they must be too massive or
too pressure-confined to be affected by feedback from young star
clusters, or the clouds could be in the outer more quiescent part of
the disk which is probed by the fiber at high galaxy inclinations.

The outflow absorption component covers a wide angle, but seems stronger
within $\sim 60\,^{\circ}$ of the disk rotation axis; it likely arises
from a large-scale galactic wind, like that seen in
Figure~\ref{Fig_M82}.  Notably, the velocity is roughly 
constant at $i < 60^{\circ}$, as expected if the outflow is radially
directed.   At $i = 60\,^{\circ}$, the EW of the disk-like and outflow 
components are roughly equal. This is consistent with our understanding of the
outflow and disk geometry: at $i < 60\,^{\circ}$, the ISM Na D
absorption is dominated by outflow, while the disk-like component
becomes important at $i > 60\,^{\circ}$. Based on these results, we
will split ``Sample A" into two 
sub-samples, ``Sample B'', with $i > 60\,^{\circ}$, and ``Sample C'' 
with  $i < 60\,^{\circ}$. There are 60,211 and 80,414 galaxies in each 
sample, respectively.

\subsection{Trends with Galaxy Physical Properties 
\label{outflow_properties}}

In this section we examine how the properties of the two Na D
absorption components (as derived from our constrained two-component
fit) vary as a function of galaxy physical properties. ``Sample B" and
``Sample C" are used to study the disk-like and outflow components,
respectively.  Our reasoning is that the absorbers in the disk 
are best studied in galaxies that are more nearly edge-on, while
absorbers in the outflow are best studied in galaxies that are more 
nearly face-on.

\begin{figure}
\bc
\hspace{-0.6cm}
\resizebox{8.5cm}{!}{\includegraphics{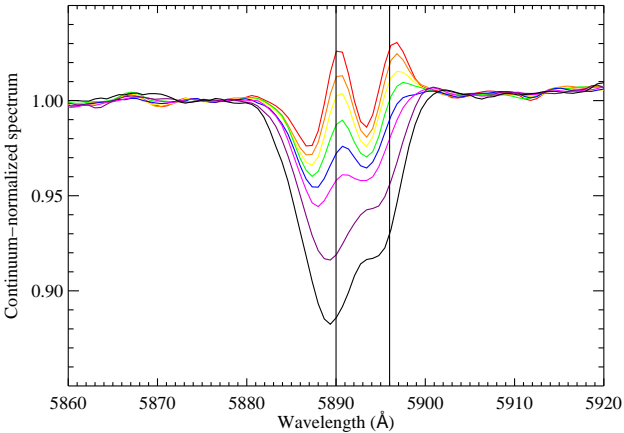}}\\%
\caption{Continuum normalized composite spectra of galaxies in 8 different 
$\Sigma_{\rm SFR}$ bins drawn from Sample C 
($i < 60^{\circ}$). As $\Sigma_{\rm SFR}$ 
increases, the absorption becomes stronger. $\Sigma_{\rm SFR}$ increases 
from red to purple.}
\label{Fig_ESFR_stack}
\ec
\end{figure}

\begin{figure*}
\bc
\hspace{-1.6cm}
\resizebox{18cm}{!}{\includegraphics{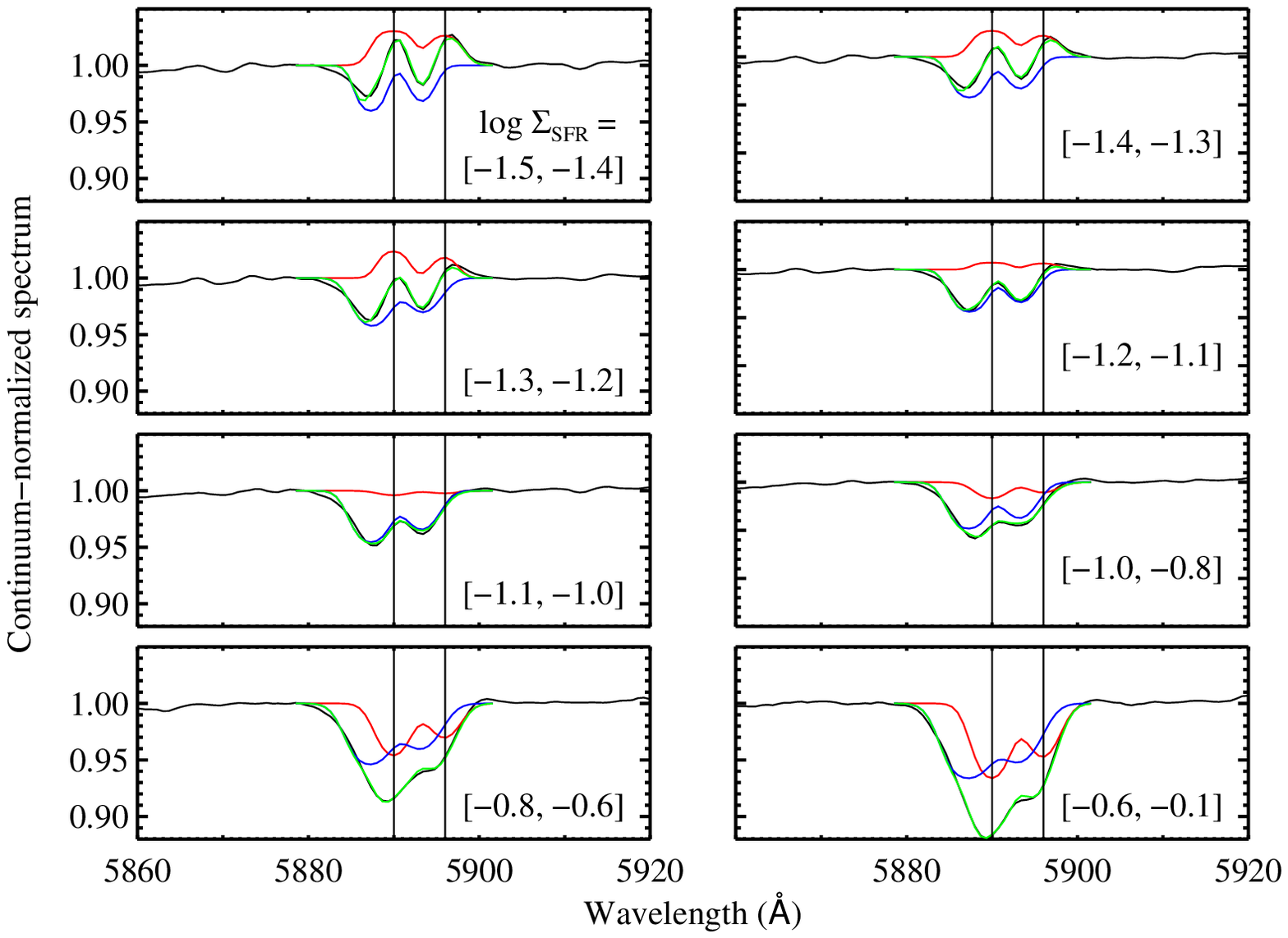}}\\%
\caption{Absorption line profile fits to composite spectra
in different $\Sigma_{\rm SFR}$ bins. (The spectra are the same as
those shown in Figure~\ref{Fig_ESFR_stack}.) The black lines show the 
continuum-normalized spectrum; the red line is 
the systemic (disk) absorption component; the blue line is the 
outflow component; and the green lines is the combined
model. The range in $\log \Sigma_{\rm SFR}$ is given
in the lower right of each panel.
\label{Fig_ESFR_fit}}
\ec
\end{figure*}

We consider each galaxy physical parameter independently and use the 
adaptive binning approach described in \S\ref{stacking} to 
construct stacked spectra.  We do not consider the full parameter
range, but rather we define a lower boundary based where
the fraction of strong Na~D absorbers (EQ $>$ 0.8~\AA) begins to rise 
steeply (see the blue lines in Fig.~\ref{Fig_frac_NaD}). 
The continuum-normalized spectra for eight different $\Sigma_{\rm SFR}$ 
bins from ``Sample C" are shown in Figure~\ref{Fig_ESFR_stack}
as an example. The strength of the absorption increases dramatically 
with increasing $\Sigma_{\rm SFR}$.  Figure~\ref{Fig_ESFR_fit} shows 
the two-component fits of these eight spectra.

\begin{figure*}
\bc
\hspace{-1.6cm}
\resizebox{18cm}{!}{\includegraphics{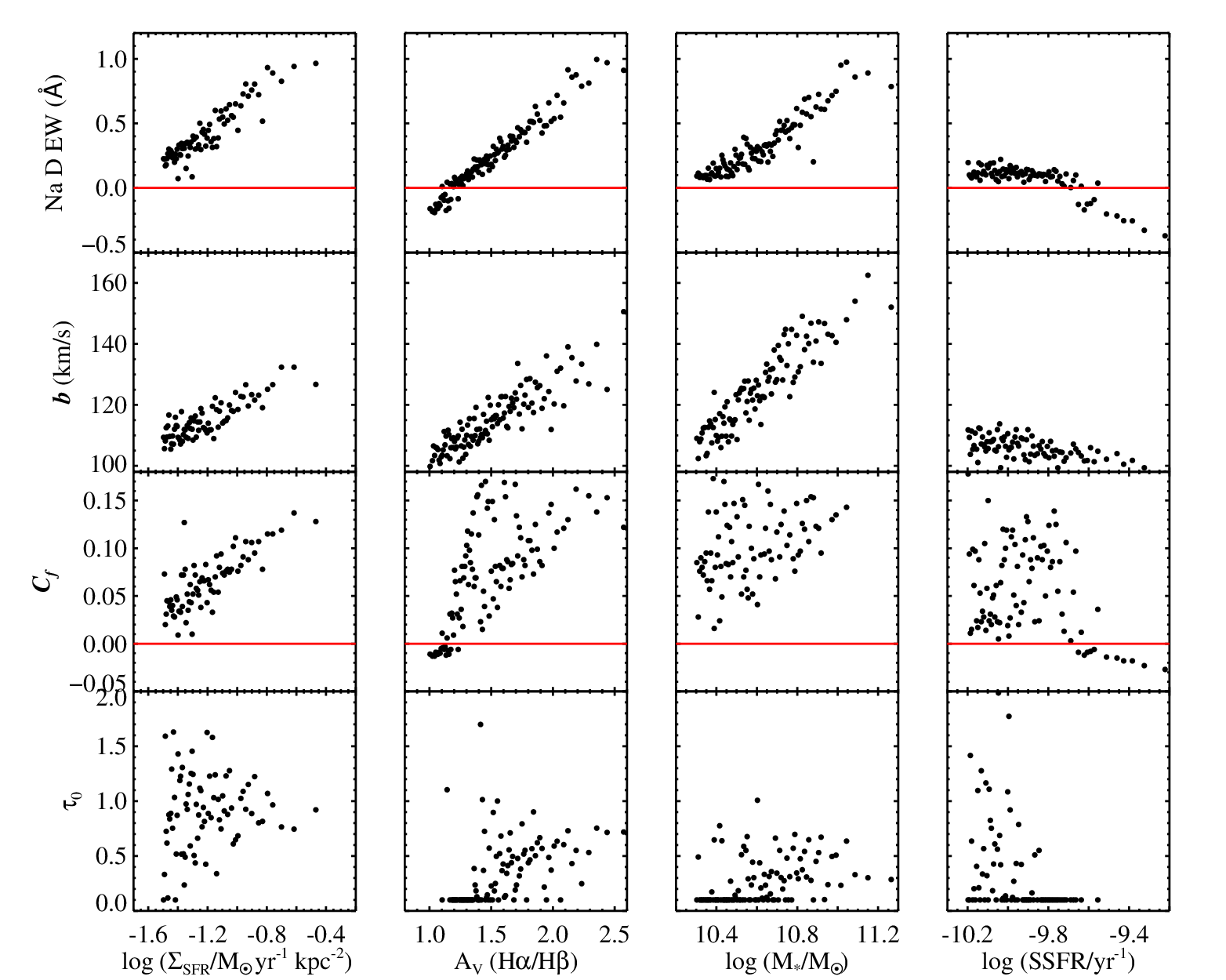}}\\%
\caption{Derived parameters of the systemic Na~D absorption component 
as a function of $\Sigma_{\rm SFR}$, $\rm A_V$, 
$M_*$, SSFR for Sample B galaixes ($i > 60^{\circ}$).
\label{Fig_NaD_sys_phys}}
\ec
\end{figure*}

\begin{figure*}
\bc
\hspace{-1.6cm}
\resizebox{18cm}{!}{\includegraphics{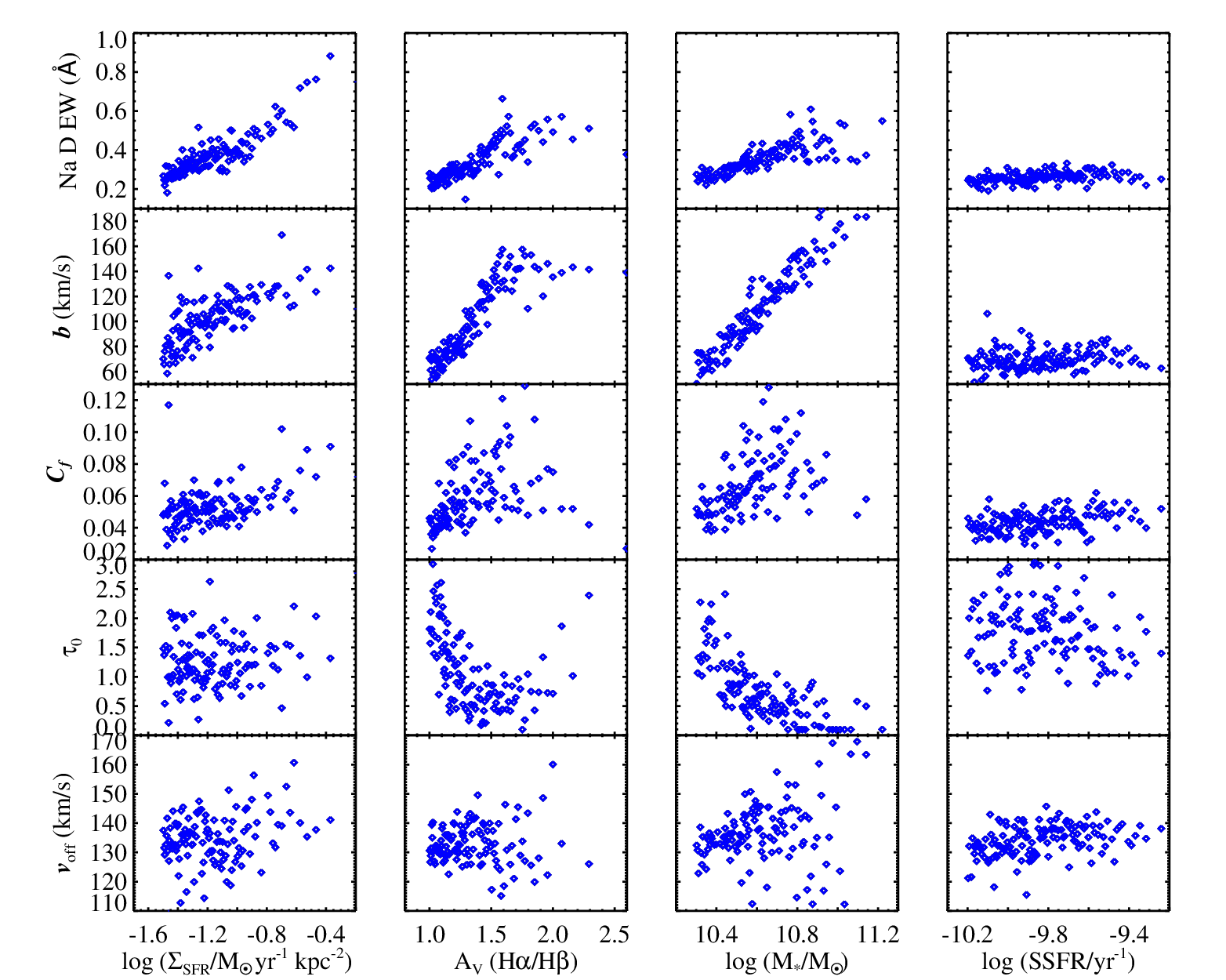}}\\%
\caption{Derived parameters of the outflow component 
as a function of $\Sigma_{\rm SFR}$, $\rm A_V$, 
$M_*$ and SSFR for Sample C galaxies ($i < 60^{\circ}$).
\label{Fig_NaD_dyn_phys}}
\ec
\end{figure*}

We plot the parameters derived from our absorption line fits ($v_{\rm
  off}$, $b$, EW, $C_f$, $\tau_0$) as a function of galaxy physical
properties ($\Sigma_{\rm SFR}$, $\rm A_V$, SSFR, $M_*$) for Samples B 
and C in Figures~\ref{Fig_NaD_sys_phys} and \ref{Fig_NaD_dyn_phys}. 
We use these plots to help discern the galaxy physical parameters that
have the greatest influence on the warm neutral medium in and outside
of galaxy disks. However, the strong correlations among the galaxy 
physical parameters illustrated in Figure~\ref{Fig_param_correlations}
make this exercise challenging. We also considered correlations with
SFR and $\Sigma_{M_*}$ (both measured in the fiber aperature). Trends 
with these parameters are very similar to those found with 
$\Sigma_{\rm SFR}$ and  $M_*$, but typically weaker and/or nosier.  For
clarity we have not included SFR and $\Sigma_{M_*}$ in our subsequent analysis.

For the systematic component (Fig.~\ref{Fig_NaD_sys_phys}), we note
the following:
\begin{itemize}
\item Na~D EW increases strongly and nearly linearly with
log($\Sigma_{\rm SFR}$), log($M_*$), and $\rm A_V$
from negative values (emission) to $\sim$1.0~\AA. We note
that the line EW is sensitive to the velocity spread of the gas, the
cloud covering factor and the optical depth.

\item The line width (which is tied to the width of the He~I line)
  also increases smoothly with all of the stacking parameters except
  SSFR.  The strongest trend seen is a 60\% increase in $b$ with
  stellar mass.  Since Sample B galaxies are highly inclined, the line
  width probably reflects the rotation speed of the inner disk. The
  apparent correlation between $b$ and $\Sigma_{\rm SFR}$, 
  $\rm A_V$ most likely results from the
  correlation of $M_{*}$ with these parameters, as show in
  Figure~\ref{Fig_param_correlations}. 

\item The covering factor changes steeply, from negative 
  values (Na~D emission) to $C_f=0.15$.  The very
  small Na~D covering factor may be related to the low filling factor
  of neutral gas in galactic disks. The trends are similar to those
  identified with the EW, but noisier due degeneracies with $\tau_0$
  exacerbated by the blended nature of our line profiles. 

\item The optical depth ranges from $\tau_0$=0 to $\sim1.5$, and it
  does not depend strongly on any of the stacking parameters. In general, 
the optical depth is the least well constrained parameter that we measure.

\end{itemize}

Examining the changes in $b$, $C_f$, and $\tau_0$, we hypothesize that
the changes in EW are driven primarily by the covering factor with an
small additional contribution due changes in the line width.  
It is difficult to determine whether $\Sigma_{\rm SFR}$, $\rm A_{v}$,
or M$_{*}$ are the primary driver of the trends. 
SSFR is clearly not an important parameter.  We address the issue of
correlations between the various physical parameters in \S\ref{primary_trends}.

In Figure \ref{Fig_NaD_dyn_phys} we explore the dependence of outflow
Na~D component on various galaxy properties.  We note the following:
\begin{itemize}
\item The Na D EW increases with $\Sigma_{\rm SFR}$, $M_*$ 
      and $\rm A_V$ (similar to the disk component).

\item The line width appears to scale strongly with $\Sigma_{\rm
    SFR}$, $M_*$ and $\rm A_V$. The tightest
  correlation appears to be with $M_*$: $b\propto M_*^{0.65}$. 

\item The covering factor is very low  ($C_f\sim0.04-0.12$) and changes by a
  factor of three, at most.  The strongest 
  correlation is with $\rm A_V$ and M$_{*}$.

\item The lines range from unsaturated to mildly saturated
  ($\tau_0\sim 3$) and $\tau_0$ displays a pronounced inverse correlation with
   $\rm A_V$, and M$_{*}$.

\item The outflow velocity does not show any strong trend with galaxy
  physical properties.  A very weak and noisy correlation is evident with 
  $\Sigma_{\rm SFR}$ and possibly with $M_*$.

\item No properties of the outflow component are well correlated with 
 the specific star formation rate.
\end{itemize}

One possible interpretation of the observed trends in $C_f$ and
$\tau_0$ is that in galaxies with low dust attenuation Na~D only
survives in very dense clouds in the outflow.  These clouds are
optically thick and comparatively rare, so the average $\tau_0$
is high and the Na~D covering factor is very low.  In galaxies with
higher dust-to-gas ratios, Na~D is shielded from ionizing radiation
in clouds with lower optical depths.  Since these clouds are more
numerous, the Na D covering factor is higher and the average optical
depth is lower.  Thus, the mean optical depth \emph{decreases}
with $\rm A_V$ while the covering factor \emph{increases}, as
observed.

The lack of trends with SSFR is somewhat surprising
since it might be expected that outflows depend on both the 
SFR and the potential well depth.  One explanation is the
strange shape of the SSFR vs. $\rm A_V$ correlation  
(Fig.\ref{Fig_param_correlations}).  Galaxies with both low and high 
SSFR have low dust attenuation. This arises because low SSFR
galaxies are predominately gas-poor massive galaxies with high
dust-to-gas ratios, while high SSFR galaxies are predominantly 
gas-rich dwarfs with low metallicities and dust-to-gas ratios.  
At intermediate values of SSFR, galaxies display a wide range 
in dust attenuation.  Clearly, dust shielding plays a key role in the
detectablity of outflows using Na~I.  

The EW of the outflow component will depend on $b$, $C_f$, and
$\tau_0$, although the influence of $\tau_0$ is minimal for galaxies  
with $\tau_0 > 1$.  The EW depends linearly on $b$ and $C_f$, but the
dynamic range probed by our sample is greater in $C_f$ (factor of
$\sim6$ vs. factor of $\sim3$.)  The EW thus primarily reflect changes
in the covering factor.  Since our measurement of the EW is 
less noisy than our measurement of $C_f$, we use the EW in examining 
tends with galaxy physical properties further below.

\subsection{Isolating the Primary Drivers of the Observed Correlations
\label{primary_trends}}

In figures \ref{Fig_NaD_sys_phys} and \ref{Fig_NaD_dyn_phys} we
identified $\rm A_V$, $\Sigma_{\rm SFR}$, and $M_*$ as important
physical parameters controlling the Na~D line profile.  However, due to
strong correlations among these parameters
(Fig.~\ref{Fig_param_correlations}), it is difficult to isolate the
primary drivers of the trends we observe.  The correlations are
sufficiently strong that it is not possible, for example, to look for
trends with $\Sigma_{\rm SFR}$ at fixed $M_{*}$. A bin with a small range
in stellar mass would also have a small range in $\Sigma_{\rm SFR}$.  To
isolate the most important physical parameters, we adopt the following
approach.  We take each bin in $M_{*}$ (for example) and further
divide it into three equal sub-bins, sorting the galaxies by
$\Sigma_{\rm SFR}$.  We are then able to examine trends between Na~D
absorption and $M_{*}$ in low, medium, and high $\Sigma_{\rm SFR}$ bins
(where the exact division between low, medium, and high changes with
$M_{*}$).  If $M_{*}$ is the primary driver of the trend, then no
difference is expected between the sub-bins.  Conversely, if 
$\Sigma_{\rm SFR}$ is the primary driver of the trend, then the three 
sub-bins will be strongly offset from one another at each value of $M_{*}$.

\begin{figure}
\bc
\hspace{-0.6cm}
\resizebox{8.5cm}{!}{\includegraphics{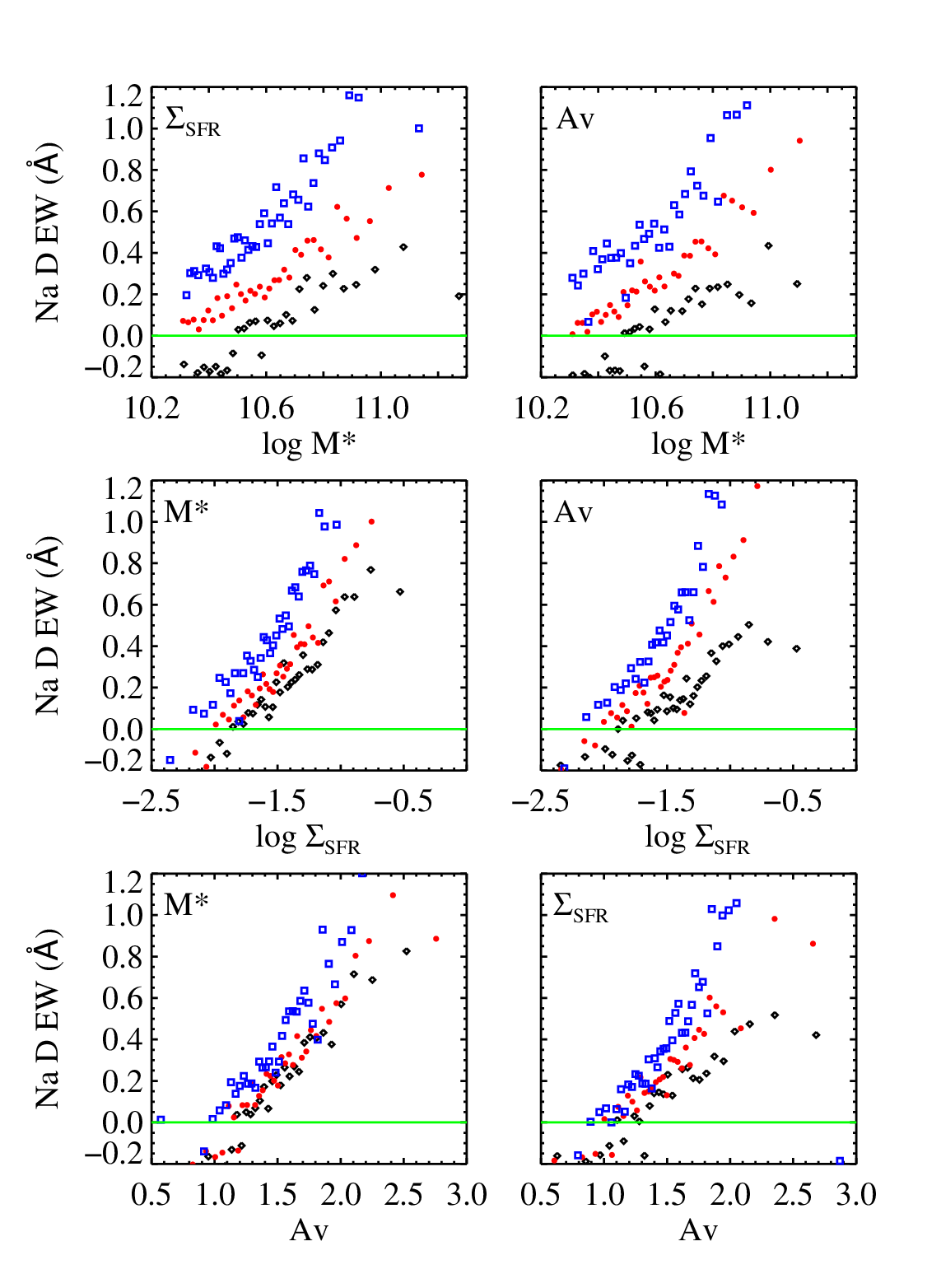}}\\%
\caption{The dependence of the Na~D EW of the systemic component on
  galaxy physical properties. In each panel, the sample is divided
  into three sub-bins by the parameter marked in the top-left corner.
  The black, red, and blue points indicate the low, medium, and high
  sub-bins respectively. This approach is designed to help mitigate
  the effects of correlations among the galaxy physical parameters
  (see \S\ref{primary_trends} for details). The physical parameters
  that play the dominant role in driving the trends will show the
  smallest offsets between the black, red, and blue points.}
\label{Fig_ew_sys}
\ec
\end{figure}

Figure~\ref{Fig_ew_sys} shows the Na~D EW of the systemic component as 
a function of $M_{*}$, $\Sigma_{\rm SFR}$, and $\rm A_V$ for galaxies in  
Sample B with log $M_*/M_\odot > 10.3$.  In each panel the sample 
is split into three sub-bins according to the galaxy parameter labeled 
in the top-left corner.  The black, red, and blue points indicate the low,
medium, and high sub-bins respectively. The dominant driver of the
observed trend (the parameter displaying the smallest
offsets between the sub-bins) appears to be $\rm A_V$.  This suggests
that high dust attenuation is important to keep Sodium in the disk
neutral. A small residual correlation is evident with $\Sigma_{\rm SFR}$
and M$_{*}$.  A correlation between $\Sigma_{\rm SFR}$ and the
Na D covering factor is expected if $\Sigma_{\rm SFR}$ scales with
the filling factor of molecular clouds, as suggested in \citet{bigiel08}.
The residual trend of EW with stellar mass is due to the larger
velocity spread of the line in disk galaxies with larger rotation speeds.

\begin{figure}
\bc
\hspace{-0.6cm}
\resizebox{8.5cm}{!}{\includegraphics{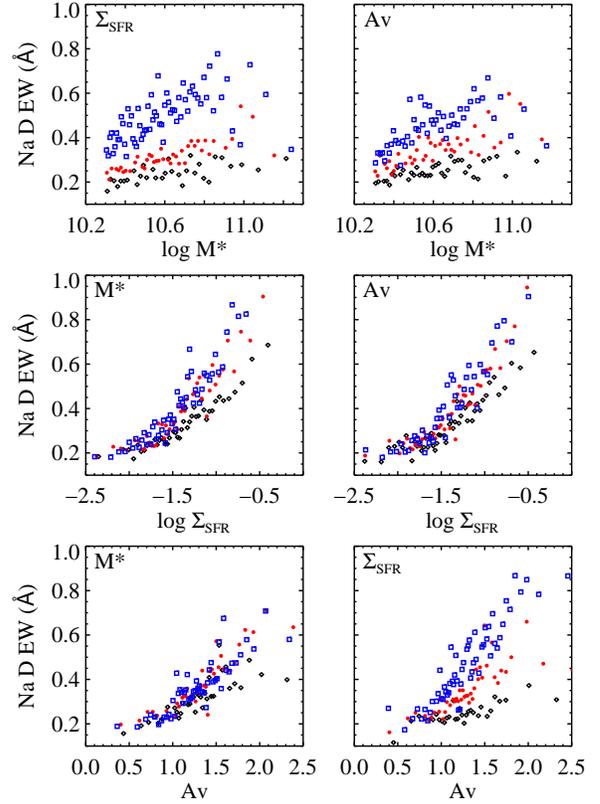}}\\%
\caption{The dependence of the Na~D EW of the outflow component on
  galaxy physical properties. As in Figure~\ref{Fig_ew_sys}, 
 the sample in each panel is divided
  into three sub-bins by the parameter marked in the top-left corner 
  (black=low, red=medium, blue=high).}
\label{Fig_ew_blue}
\ec
\end{figure}

Figure~\ref{Fig_ew_blue} shows the Na~D EW of the outflow component as
a function of $M_{*}$, $\Sigma_{\rm SFR}$, and $\rm A_V$ for galaxies in
Sample C with log($M_*/M_\odot) > 10.3$.  As in
Figure~\ref{Fig_ew_sys}, the galaxies are split into three sub-bins
according to various physical properties.  The plots with stellar mass
on the x-axis show very large discrepancies between high and low
sub-bins in $\Sigma_{\rm SFR}$ and $\rm A_V$, therefore, we infer that
$M_{*}$ is the least important physical parameter in determining the
EW.  The most important parameter influencing the line EW appears to
be $\Sigma_{\rm SFR}$ followed closely by $\rm A_V$.  The physical
explanation for these correlations is straightforward: star formation
surface density determines the amount of material lofted above the
disk and the dust attenuation influences the survival of Na I in the
clouds.

A similar analysis applied to the outflow velocity ($v_{\rm off}$)
instead of the EW does not yield any insight into the dominant
physical parameter determining the outflow speed.  We note that the
dynamic range in $v_{\rm off}$ is very small (120 - 160 km~s$^{-1}$),
and this makes identification of trends particularly challenging.  It
is also important to keep in mind that the range in galaxy physical
properties probed by our study is quite small -- typically of order 1
dex.  For instance, we are looking for trends over a range in stellar
mass of $\log(M_*/M_{\sun}) = 10.3-11.3$, whereas the full galaxy
population spans over six orders of magnitude in stellar mass.  This
limitation is imposed by our need to study objects where Na~D
absorption is common (Fig.~\ref{Fig_frac_NaD}).

In spite of these limitations, we
can use our results to investigate a basic physical model of the
outflow.  Following \citet{martin09} we suggest that the velocity of
the line centroid reflects the speed of the swept-up shell of 
interstellar gas at the point where it blows out of the disk.  
Combining equations 3 and 6 from \citet{strickland04}, for the
velocity of a shell of gas at blow-out, we find
$v_{\rm off} \propto (\Sigma_{\rm SFR}/\rho_0)^{1/3}$ where $\rho_0$ is the
gas density.  Assuming $\rho_0 \propto \Sigma_{gas}/H_z$, where
$H_z$ is the scale height of the disk, and 
$\Sigma_{gas} \propto \Sigma_{\rm SFR}^{1/1.4}$ \citep{Kennicutt98b},
we find  $v_{\rm off} \propto \Sigma_{\rm SFR}^{0.1} H_{z}^{1/3}$.  This is
consistent with the very shallow scaling we see between outflow
velocity and $\Sigma_{\rm SFR}$.

\begin{figure}
\bc
\hspace{-0.6cm}
\resizebox{8.5cm}{!}{\includegraphics{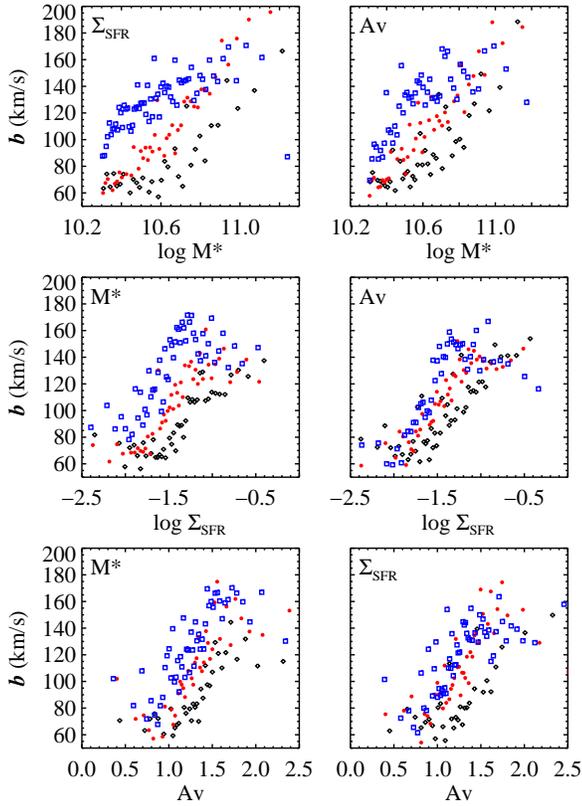}}\\%
\caption{The dependence of the line width of the outflow component on
  galaxy physical properties. As in Figure~\ref{Fig_ew_sys}, 
 the sample in each panel is divided
  into three sub-bins by the parameter marked in the top-left corner 
  (black=low, red=medium, blue=high).
\label{Fig_b_blue}}
\ec
\end{figure}

Once the cool entrained gas has escaped the disk, it may be accelerated
further by the ram pressure of the hot wind, or by radiation pressure
on dust grains. This halo gas likely contributes to the wings of the
line profile \citep{martin09}.  In Figure~\ref{Fig_b_blue}, we
examine the physical drivers of the line width of the Na~D outflow
component.  We see that the line width is sensitive to all three parameters:
$M_{*}, \Sigma_{\rm SFR}$, and $\rm A_V$.  The latter two are intuitive:
cloud acceleration is likely related to $\Sigma_{\rm SFR}$, while dust
attenuation is important to keep the gas neutral.
The positive correlation between line width and stellar mass is more puzzling.  
It may be that clouds with velocities above the halo escape velocity
quickly dissipate after leaving the pressure-confining hot medium of
the halo. The larger line widths in more massive galaxies may
therefore reflect the larger size of the hot halo.

\subsection{Comparison with Previous Work}
\begin{figure}
\bc
\hspace{-0.6cm}
\resizebox{8.5cm}{!}{\includegraphics{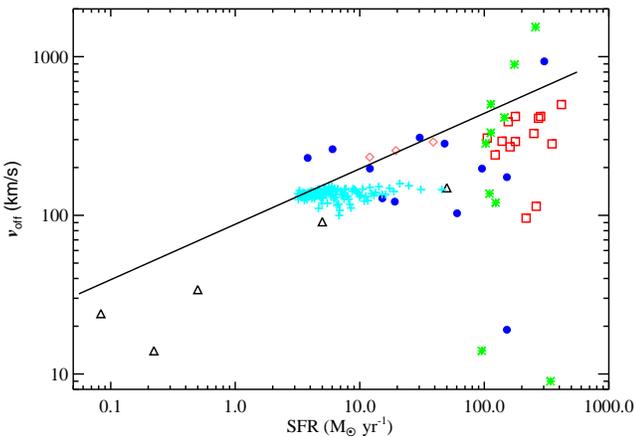}}\\%
\caption{Na D outflow velocity vs. galactic SFR.
Our data is shown as cyan crosses; green asterisks are 1 Jy ULIRGs
from \citet{rupke02}; red squares are ULIRGs from \citet{martin05}; 
blue circles are LIRGs from \citet{heckman00}; black triangles are 
dwarf starbursts from \citet{schwartz04}, and the pink diamonds are 
from \citet{weiner09}. The black line is the upper envelope fit given by
\citet{martin05}.
\label{fig_lit_compare}
}
\ec
\end{figure}

A correlation between outflow velocity and galaxy SFR and
rotation speed was identified by \citet{rupke05b} and
\citet{martin05} in starburst galaxies. In
Figure~\ref{fig_lit_compare} we compare our data (cyan crosses) to
these results.  For consistency with the literature data, we use
global (i.e., aperture-corrected) SFRs.  Green asterisks show 1 Jy
ULIRGs from \citet{rupke02}; red squares indicate ULIRGs from
\citet{martin05}; blue circles are LIRGs from \citet{heckman00}; 
pink diamonds are $z\sim1.4$ galaxies from \citep{weiner09}; and
black triangles are dwarf starbursts from \citet{schwartz04}. All 
the data are corrected to a Kroupa IMF.  

The velocities of our composite spectra show good agreement with the
velocities derived from individual galaxies in the same SFR range
\citep{heckman00, schwartz04}.  However, the trend with SFR is weaker
than the correlation suggested by \citet[][black line]{martin05}. We
note that slope of this correlation is strongly influenced by three
dwarf galaxies with very low SFRs. The large scatter on $v_{\rm off}$
at fixed SFR for the individual measurements may be due both to 
intrinsic variations and inclination effects.

Based on a sample of high stellar mass galaxies between 
$z \sim 1.4$ and $z \sim 1$, \citet{rubin09} suggest that outflow 
absorption strength (measured from Mg~II $\lambda\lambda2796,2803$)
is more closely associated with SFR than with $M_*$, and it does not 
increase with increasing SSFR. Although they have not detected a strong 
dependence of outflow absorption strength on $\Sigma_{\rm SFR}$, they 
have by no means ruled it out due to the spectrum S/N limit.

\section{Summary}

In this paper, we study interstellar Na~I ``D''
$\lambda\lambda5890,5896$ absorption in a large sample
of star forming galaxies drawn from SDSS DR7.  At the low SFRs probed
by our sample, cool stars with prominent stellar Na D lines make a
significant contribution to the integrated light.  We account for this
by modeling the full galaxy continuum using the CB08 stellar
population synthesis models.  We find a high incidence of strong ISM
Na D absorption lines (EW $> 0.8$~\AA) in galaxies that are massive,
heavily dust attenuated, and that have high $\Sigma_{\rm SFR}$(Fig.~\ref{Fig_frac_NaD}.)

We use the spectral stacking technique to increase the signal-to-noise
ratio of the spectra and the completeness of the sample and stack
spectra in various bins of galaxy physical properties.  In the
continuum-normalized spectra we identify two interstellar Na D
absorption components --- one at the systemic velocity, which is most
pronounced in edge-on galaxies, and one outflow (blue-shifted) component which
is dominant in face-on galaxies.  We use two-component absorption line
profile fits to measure the outflow velocity, the absorption line EW,
the line width, covering factor, and the optical depth.
As highlighted in the appendix, our results are robust with respect 
to the modeling of the stellar continuum. We find that:
\begin{itemize}

\item The ISM Na D absorption arises from cool gas in the disk,
  and from cool gas entrained in a galactic wind that is  
  perpendicular to the disk and has an opening 
  angle of $\sim60\,^{\circ}$.   

\item For the systemic (disk) component, the Na D EW depends 
most strongly on the dust attenuation.  Dust attenuation is 
undoubtedly important in shielding Na~I from the ionizing 
radiation of OB stars.  At fixed
$\rm A_V$, there is an additional dependence on $\Sigma_{\rm
SFR}$.  Galaxies with higher $\Sigma_{\rm SFR}$ probably have a
higher filling factor of cold clouds and thus a higher Na~D covering 
fraction.

\item For the outflow component the Na D EW depends strongly
on $\Sigma_{\rm SFR}$ and secondarily on $\rm A_V$.   We hypothesize
that the star formation surface density determines the amount of 
material lofted above the disk, and the dust attenuation 
influences the survival of Na I in the clouds.   

\item The covering factor of Na~I in the outflow increases with
increasing dust attenuation while the optical depth of the
absorption decreases.  This suggests that Na D is able to survive in
clouds with lower column density as the dust-to-gas ratio increases. 

\item The outflow velocity (line centroid) does not depend strongly on
  any galaxy physical parameters over the limited dynamic 
  range in physical properties probed by our study.  There is some
  evidence for a very shallow trend, 
  $v_{\rm off} \propto \Sigma_{\rm SFR}^{0.1}$, which is consistent
  with theoretical expectations for the velocity of a swept-up 
  shell of gas at the point where it blows out of the disk.  

\item The line width of the outflow component is sensitive to  
$\Sigma_{\rm SFR}$, $\rm A_V$, and $M_{*}$.  These parameters are 
likely to influence the acceleration of halo clouds (by ram or 
radiation pressure), the ability of clouds to remain neutral, and 
the length of time a cloud can remain pressure confined.

\end{itemize}


The question of how galactic scale outflows are influenced by AGNs is also of
great interest. In a companion paper, we will study the Na D
absorption properties in SDSS galaxies hosting type~2 AGNs using methods
similar to those described here.

  The Na~D ISM absorption line provides a convenient probe of galactic
  winds due to its location in the optical. However, the low EW width
  of the line in typical galaxy spectra and the strength of the Na~D
  stellar feature make interstellar Na~D absorption a challenge to
  measure accurately.  The strong sensitivity of Na~I to ionization by
  OB stars is also a very serious concern, in particular for studies
  that aim to compute mass outflow rates from Na~I column densities
  and velocities. Higher redshift studies that make use of higher
  ionization transitions such as Mg~II~$\lambda2796, 2803$ may
  ultimately be able to provide more robust constraints over a larger
  range in galaxy physical properties.

  Theoretical models of galaxy evolution have begun to incorporate
  galactic winds, using various ad-hoc prescriptions based on our
  knowledge of the cool gas. However, the majority the energy and
  newly-synthesized metals in the outflow resides in the hot X-ray
  emitting phase of the wind (D.~K. Strickland \& D.~C. Dinge, 2010, in prep.).  To
  truly understand the impact of galactic winds on the evolution of
  galaxies and the intergalactic medium, it is crucial to measure the
  chemical composition and velocity of the hot gas. This will require
  a high sensitivity X-ray imaging spectrometer such as the planned
  {\it International X-ray Observatory.}

\acknowledgements{We are very grateful to the referee for useful 
suggestions. C.A.T thanks the Alexander von Humboldt foundation for their generous
support while much of this work was completed.

Funding for the SDSS and SDSS-II has been provided by the Alfred P. Sloan 
Foundation, the Participating Institutions, the National Science Foundation, 
the U.S. Department of Energy, the National Aeronautics and Space 
Administration, the Japanese Monbukagakusho, the Max Planck Society, and 
the Higher Education Funding Council for England. 
The SDSS Web Site is http://www.sdss.org/.

The SDSS is managed by the Astrophysical Research Consortium for 
the Participating Institutions. The Participating Institutions are 
the American Museum of Natural History, Astrophysical Institute Potsdam, 
University of Basel, University of Cambridge, Case Western Reserve 
University, University of Chicago, Drexel University, Fermilab, the 
Institute for Advanced Study, the Japan Participation Group, Johns Hopkins 
University, the Joint Institute for Nuclear Astrophysics, the Kavli 
Institute for Particle Astrophysics and Cosmology, the Korean Scientist 
Group, the Chinese Academy of Sciences (LAMOST), Los Alamos National 
Laboratory, the Max-Planck-Institute for Astronomy (MPIA), the 
Max-Planck-Institute for Astrophysics (MPA), New Mexico State University, 
Ohio State University, University of Pittsburgh, University of Portsmouth, 
Princeton University, the United States Naval Observatory, and the 
University of Washington.}


\begin{thebibliography}{}

\bibitem[\protect\citeauthoryear{{Abazajian}, {Adelman-McCarthy},
  {Ag{\"u}eros}, {Allam}, {Anderson}, {Anderson}, {Annis} \&
  {Bahcall}}{{Abazajian} et~al.}{2004}]{abazajian04}
{Abazajian} K.,  {Adelman-McCarthy} J.~K.,  {Ag{\"u}eros} M.~A.,  {Allam}
  S.~S.,  {Anderson} K.,  {Anderson} S.~F.,  {Annis} J.,    {Bahcall} N.~A.,
  et al., 2004, \aj, 128, 502

\bibitem[\protect\citeauthoryear{{Abazajian}, {Adelman-McCarthy},
  {Ag{\"u}eros}, {Allam}, {Allende Prieto}, {An}, {Anderson} \&
  {Anderson}}{{Abazajian} et~al.}{2009}]{abazajian09}
{Abazajian} K.~N.,  {Adelman-McCarthy} J.~K.,  {Ag{\"u}eros} M.~A.,  {Allam}
  S.~S.,  {Allende Prieto} C.,  {An} D.,  {Anderson} K.~S.~J.,    {Anderson}
  S.~F.,  et al., 2009, \apjs, 182, 543

\bibitem[\protect\citeauthoryear{{Adelman-McCarthy}, {Ag{\"u}eros}, {Allam},
  {Allende Prieto}, {Anderson}, {Anderson}, {Annis} \&
  {Bahcall}}{{Adelman-McCarthy} et~al.}{2008}]{adelman08}
{Adelman-McCarthy} J.~K.,  {Ag{\"u}eros} M.~A.,  {Allam} S.~S.,  {Allende
  Prieto} C.,  {Anderson} K.~S.~J.,  {Anderson} S.~F.,  {Annis} J.,
  {Bahcall} N.~A.,  et al., 2008, \apjs, 175, 297

\bibitem[\protect\citeauthoryear{{Benson}, {Bower}, {Frenk}, {Lacey}, {Baugh}
  \& {Cole}}{{Benson} et~al.}{2003}]{Benson03}
{Benson} A.~J.,  {Bower} R.~G.,  {Frenk} C.~S.,  {Lacey} C.~G.,  {Baugh} C.~M.,
     {Cole} S.,  2003, \apj, 599, 38

\bibitem[\protect\citeauthoryear{{Bica}, {Pastoriza}, {da Silva}, {Dottori} \&
  {Maia}}{{Bica} et~al.}{1991}]{bica91}
{Bica} E.,  {Pastoriza} M.~G.,  {da Silva} L.~A.~L.,  {Dottori} H.,    {Maia}
  M.,  1991, \aj, 102, 1702

\bibitem[\protect\citeauthoryear{{Bigiel}, {Leroy}, {Walter}, {Brinks}, {de
  Blok}, {Madore} \& {Thornley}}{{Bigiel} et~al.}{2008}]{bigiel08}
{Bigiel} F.,  {Leroy} A.,  {Walter} F.,  {Brinks} E.,  {de Blok} W.~J.~G.,
  {Madore} B.,    {Thornley} M.~D.,  2008, \aj, 136, 2846

\bibitem[\protect\citeauthoryear{{Bregman}}{{Bregman}}{1980}]{bregman80}
{Bregman} J.~N.,  1980, \apj, 236, 577

\bibitem[\protect\citeauthoryear{{Brinchmann}, {Charlot}, {White}, {Tremonti},
  {Kauffmann}, {Heckman} \& {Brinkmann}}{{Brinchmann}
  et~al.}{2004}]{brinchmann04}
{Brinchmann} J.,  {Charlot} S.,  {White} S.~D.~M.,  {Tremonti} C.,  {Kauffmann}
  G.,  {Heckman} T.,    {Brinkmann} J.,  2004, \mnras, 351, 1151

\bibitem[\protect\citeauthoryear{{Calzetti}, {Kinney} \&
  {Storchi-Bergmann}}{{Calzetti} et~al.}{1994}]{Calzetti94}
{Calzetti} D.,  {Kinney} A.~L.,    {Storchi-Bergmann} T.,  1994, \apj, 429, 582

\bibitem[\protect\citeauthoryear{{Cardelli}, {Clayton} \& {Mathis}}{{Cardelli}
  et~al.}{1989}]{Cardelli89}
{Cardelli} J.~A.,  {Clayton} G.~C.,    {Mathis} J.~S.,  1989, \apj, 345, 245

\bibitem[\protect\citeauthoryear{{Chevalier} \& {Clegg}}{{Chevalier} \&
  {Clegg}}{1985}]{chevalier85}
{Chevalier} R.~A.,  {Clegg} A.~W.,  1985, \nat, 317, 44

\bibitem[\protect\citeauthoryear{{Cole}, {Lacey}, {Baugh} \& {Frenk}}{{Cole}
  et~al.}{2000}]{Cole00}
{Cole} S.,  {Lacey} C.~G.,  {Baugh} C.~M.,    {Frenk} C.~S.,  2000, \mnras,
  319, 168

\bibitem[\protect\citeauthoryear{{Dahlem}, {Weaver} \& {Heckman}}{{Dahlem}
  et~al.}{1998}]{dahlem98}
{Dahlem} M.,  {Weaver} K.~A.,    {Heckman} T.~M.,  1998, \apjs, 118, 401

\bibitem[\protect\citeauthoryear{{Dalcanton}}{{Dalcanton}}{2007}]{dalcanton07}
{Dalcanton} J.~J.,  2007, \apj, 658, 941

\bibitem[\protect\citeauthoryear{{de Vaucouleurs}}{{de
  Vaucouleurs}}{1948}]{devau48}
{de Vaucouleurs} G.,  1948, Annales d'Astrophysique, 11, 247

\bibitem[\protect\citeauthoryear{{Engelbracht}, {Kundurthy}, {Gordon}, {Rieke},
  {Kennicutt}, {Smith}, {Regan}, {Makovoz}, {Sosey}, {Draine}, {Helou},
  {Armus}, {Calzetti}, {Meyer}, {Murphy}, {Dale}, {Buckalew} \&
  {Sheth}}{{Engelbracht} et~al.}{2006}]{engel06}
{Engelbracht} C.~W.,  {Kundurthy} P.,  {Gordon} K.~D.,  {Rieke} G.~H.,
  {Kennicutt} R.~C.,  {Smith} J.,  {Regan} M.~W.,  {Makovoz} D.,  et al.,  2006,
  \apjl, 642, L127

\bibitem[\protect\citeauthoryear{{Fujita}, {Martin}, {Low}, {New} \&
  {Weaver}}{{Fujita} et~al.}{2009}]{fujita09}
{Fujita} A.,  {Martin} C.~L.,  {Low} M.,  {New} K.~C.~B.,    {Weaver} R.,
  2009, \apj, 698, 693

\bibitem[\protect\citeauthoryear{{Fukugita}, {Ichikawa}, {Gunn}, {Doi},
  {Shimasaku} \& {Schneider}}{{Fukugita} et~al.}{1996}]{fukugita96}
{Fukugita} M.,  {Ichikawa} T.,  {Gunn} J.~E.,  {Doi} M.,  {Shimasaku} K.,
  {Schneider} D.~P.,  1996, \aj, 111, 1748

\bibitem[\protect\citeauthoryear{{Garnett}}{{Garnett}}{2002}]{garnett02}
{Garnett} D.~R.,  2002, \apj, 581, 1019

\bibitem[\protect\citeauthoryear{{Gonz{\'a}lez Delgado}, {Cervi{\~n}o},
  {Martins}, {Leitherer} \& {Hauschildt}}{{Gonz{\'a}lez Delgado}
  et~al.}{2005}]{gonzalez05}
{Gonz{\'a}lez Delgado} R.~M.,  {Cervi{\~n}o} M.,  {Martins} L.~P.,  {Leitherer}
  C.,    {Hauschildt} P.~H.,  2005, \mnras, 357, 945

\bibitem[\protect\citeauthoryear{{Gunn}, {Carr}, {Rockosi}, {Sekiguchi},
  {Berry}, {Elms}, {de Haas} \& {Ivezi{\'c}}}{{Gunn} et~al.}{1998}]{gunn98}
{Gunn} J.~E.,  {Carr} M.,  {Rockosi} C.,  {Sekiguchi} M.,  {Berry} K.,  {Elms}
  B.,  {de Haas} E.,    {Ivezi{\'c}} {\v Z}.,  et al., 1998, \aj, 116, 3040

\bibitem[\protect\citeauthoryear{{Heckman}}{{Heckman}}{1980}]{heckman80}
{Heckman} T.~M.,  1980, \aap, 87, 142

\bibitem[\protect\citeauthoryear{{Heckman}}{{Heckman}}{2002}]{heckman02}
{Heckman} T.~M.,  2002, in {J.~S.~Mulchaey \& J.~T.~Stocke} ed., Extragalactic
  Gas at Low Redshift Vol.~254 of Astronomical Society of the Pacific
  Conference Series, {Galactic Superwinds Circa 2001}.
pp 292--+

\bibitem[\protect\citeauthoryear{{Heckman}, {Armus} \& {Miley}}{{Heckman}
  et~al.}{1990}]{heckman90}
{Heckman} T.~M.,  {Armus} L.,    {Miley} G.~K.,  1990, \apjs, 74, 833

\bibitem[\protect\citeauthoryear{{Heckman}, {Lehnert}, {Strickland} \&
  {Armus}}{{Heckman} et~al.}{2000}]{heckman00}
{Heckman} T.~M.,  {Lehnert} M.~D.,  {Strickland} D.~K.,    {Armus} L.,  2000,
  \apjs, 129, 493

\bibitem[\protect\citeauthoryear{{Hopkins}, {Hernquist}, {Cox}, {Robertson} \&
  {Springel}}{{Hopkins} et~al.}{2006}]{Hopkins06}
{Hopkins} P.~F.,  {Hernquist} L.,  {Cox} T.~J.,  {Robertson} B.,    {Springel}
  V.,  2006, \apjs, 163, 50

\bibitem[\protect\citeauthoryear{{Jacoby}, {Hunter} \& {Christian}}{{Jacoby}
  et~al.}{1984}]{jacoby84}
{Jacoby} G.~H.,  {Hunter} D.~A.,    {Christian} C.~A.,  1984, \apjs, 56, 257

\bibitem[\protect\citeauthoryear{{Jenkins}}{{Jenkins}}{1986}]{Jenkins86}
{Jenkins} E.~B.,  1986, \apj, 304, 739

\bibitem[\protect\citeauthoryear{{Kahn}}{{Kahn}}{1981}]{kahn81}
{Kahn} F.~D.,  1981, in {F.~D.~Kahn} ed., Investigating the Universe Vol.~91 of
  Astrophysics and Space Science Library, {Dynamics of the galactic fountain}.
pp 1--28

\bibitem[\protect\citeauthoryear{{Kauffmann}}{{Kauffmann}}{2003a}]{kauffmann03a}
{Kauffmann} G., {Heckman} T.~M.,  {White} S.~D.~M.,  {Charlot} S.,  {Tremonti}
  C.,  {Brinchmann} J.,  {Bruzual} G.,    {Peng} E.~W., et al., 2003a, \mnras, 341, 33

\bibitem[\protect\citeauthoryear{{Kauffmann}}{{Kauffmann}}{2003b}]{kauffmann03%
b}
{Kauffmann} G., {Heckman} T.~M.,  {White} S.~D.~M.,  {Charlot} S.,  {Tremonti}
  C.,  {Peng} E.~W.,  {Seibert} M.,    {Brinkmann} J., et al., 2003b, \mnras, 341, 54

\bibitem[\protect\citeauthoryear{{Kauffmann}}{{Kauffmann}}{2003c}]{kauffmann03%
c}
{Kauffmann} G., {Heckman} T.~M.,  {Tremonti} C.,  {Brinchmann} J.,  {Charlot}
  S.,  {White} S.~D.~M.,  {Ridgway} S.~E.,    {Brinkmann} J., et al., 2003c, \mnras, 346, 1055

\bibitem[\protect\citeauthoryear{{Kennicutt}
  Jr.}{{Kennicutt}}{1998a}]{kennicutt98a}
{Kennicutt} Jr. R.~C.,  1998a, \araa, 36, 189

\bibitem[\protect\citeauthoryear{{Kennicutt}
  Jr.}{{Kennicutt}}{1998b}]{Kennicutt98b}
{Kennicutt} Jr. R.~C.,  1998b, \apj, 498, 541

\bibitem[\protect\citeauthoryear{{Kroupa}}{{Kroupa}}{2001}]{Kroupa01}
{Kroupa} P.,  2001, \mnras, 322, 231

\bibitem[\protect\citeauthoryear{{Lehnert} \& {Heckman}}{{Lehnert} \&
  {Heckman}}{1996}]{lehnert96}
{Lehnert} M.~D.,  {Heckman} T.~M.,  1996, \apj, 462, 651

\bibitem[\protect\citeauthoryear{{Lupton}, {Gunn}, {Ivezi{\'c}}, {Knapp} \&
  {Kent}}{{Lupton} et~al.}{2001}]{lupton01}
{Lupton} R.,  {Gunn} J.~E.,  {Ivezi{\'c}} Z.,  {Knapp} G.~R.,    {Kent} S.,
  2001, in {F.~R.~Harnden Jr., F.~A.~Primini, \& H.~E.~Payne} ed., Astronomical
  Data Analysis Software and Systems X Vol.~238 of Astronomical Society of the
  Pacific Conference Series, {The SDSS Imaging Pipelines}.
pp 269--+

\bibitem[\protect\citeauthoryear{{Martin}}{{Martin}}{2005}]{martin05}
{Martin} C.~L.,  2005, \apj, 621, 227

\bibitem[\protect\citeauthoryear{{Martin}}{{Martin}}{2006}]{martin06}
{Martin} C.~L.,  2006, \apj, 647, 222

\bibitem[\protect\citeauthoryear{{Martin} \& {Bouch{\'e}}}{{Martin} \&
  {Bouch{\'e}}}{2009}]{martin09}
{Martin} C.~L.,  {Bouch{\'e}} N.,  2009, \apj, 703, 1394

\bibitem[\protect\citeauthoryear{{Martin}, {Kobulnicky} \& {Heckman}}{{Martin}
  et~al.}{2002}]{martin02}
{Martin} C.~L.,  {Kobulnicky} H.~A.,    {Heckman} T.~M.,  2002, \apj, 574, 663

\bibitem[\protect\citeauthoryear{{M{\'e}nard}, {Wild}, {Nestor}, {Quider} \&
  {Zibetti}}{{M{\'e}nard} et~al.}{2009}]{menard09}
{M{\'e}nard} B.,  {Wild} V.,  {Nestor} D.,  {Quider} A.,    {Zibetti} S.,
  2009, ArXiv: 0912.3263

\bibitem[\protect\citeauthoryear{{Murray}, {Martin}, {Quataert} \&
  {Thompson}}{{Murray} et~al.}{2007}]{murray07}
{Murray} N.,  {Martin} C.~L.,  {Quataert} E.,    {Thompson} T.~A.,  2007, \apj,
  660, 211

\bibitem[\protect\citeauthoryear{{Murray}, {Quataert} \& {Thompson}}{{Murray}
  et~al.}{2005}]{murray05}
{Murray} N.,  {Quataert} E.,    {Thompson} T.~A.,  2005, \apj, 618, 569

\bibitem[\protect\citeauthoryear{{Mutchler}, {Bond}, {Christian}, {Frattare},
  {Hamilton}, {Januszewski}, {Levay}, {Mountain}, {Noll}, {Royle}, {Gallagher}
  \& {Puxley}}{{Mutchler} et~al.}{2007}]{mutchler07}
{Mutchler} M.,  {Bond} H.~E.,  {Christian} C.~A.,  {Frattare} L.~M.,
  {Hamilton} F.,  {Januszewski} W.,  {Levay} Z.~G.,  {Mountain} M.,  et al.,  
  2007, \pasp, 119, 1

\bibitem[\protect\citeauthoryear{{Ohyama}, {Taniguchi}, {Iye}, {Yoshida},
  {Sekiguchi}, {Takata}, {Saito} \& {Kawabata}}{{Ohyama}
  et~al.}{2002}]{ohyama02}
{Ohyama} Y.,  {Taniguchi} Y.,  {Iye} M.,  {Yoshida} M.,  {Sekiguchi} K.,
  {Takata} T.,  {Saito} Y.,    {Kawabata} K.~S. et al.,  2002, \pasj, 54, 891

\bibitem[\protect\citeauthoryear{{Oppenheimer}, {Dav{\'e}}, {Kere{\v s}},
  {Fardal}, {Katz}, {Kollmeier} \& {Weinberg}}{{Oppenheimer}
  et~al.}{2009}]{Oppenheimer09}
{Oppenheimer} B.~D.,  {Dav{\'e}} R.,  {Kere{\v s}} D.,  {Fardal} M.,  {Katz}
  N.,  {Kollmeier} J.~A.,    {Weinberg} D.~H.,  2009, ArXiv: 0912.0519

\bibitem[\protect\citeauthoryear{{Osterbrock} \& {Ferland}}{{Osterbrock} \&
  {Ferland}}{2006}]{oster06}
{Osterbrock} D.~E.,  {Ferland} G.~J.,  2006, {Astrophysics of gaseous nebulae
  and active galactic nuclei}

\bibitem[\protect\citeauthoryear{{Padilla} \& {Strauss}}{{Padilla} \&
  {Strauss}}{2008}]{padilla08}
{Padilla} N.~D.,  {Strauss} M.~A.,  2008, \mnras, 388, 1321

\bibitem[\protect\citeauthoryear{{Pettini}, {Rix}, {Steidel}, {Adelberger},
  {Hunt} \& {Shapley}}{{Pettini} et~al.}{2002}]{pettini02}
{Pettini} M.,  {Rix} S.~A.,  {Steidel} C.~C.,  {Adelberger} K.~L.,  {Hunt}
  M.~P.,    {Shapley} A.~E.,  2002, \apj, 569, 742

\bibitem[\protect\citeauthoryear{{Pettini}, {Steidel}, {Adelberger},
  {Dickinson} \& {Giavalisco}}{{Pettini} et~al.}{2000}]{pettini00}
{Pettini} M.,  {Steidel} C.~C.,  {Adelberger} K.~L.,  {Dickinson} M.,
  {Giavalisco} M.,  2000, \apj, 528, 96

\bibitem[\protect\citeauthoryear{{Phillips}}{{Phillips}}{1993}]{phillips93}
{Phillips} A.~C.,  1993, \aj, 105, 486

\bibitem[\protect\citeauthoryear{{Rubin}, {Weiner}, {Koo}, {Martin},
  {Prochaska}, {Coil} \& {Newman}}{{Rubin} et~al.}{2009}]{rubin09}
{Rubin} K.~H.~R.,  {Weiner} B.~J.,  {Koo} D.~C.,  {Martin} C.~L.,  {Prochaska}
  J.~X.,  {Coil} A.~L.,    {Newman} J.~A.,  2009, ArXiv: 0912.2343

\bibitem[\protect\citeauthoryear{{Rupke}, {Veilleux} \& {Sanders}}{{Rupke}
  et~al.}{2002}]{rupke02}
{Rupke} D.~S.,  {Veilleux} S.,    {Sanders} D.~B.,  2002, \apj, 570, 588

\bibitem[\protect\citeauthoryear{{Rupke}, {Veilleux} \& {Sanders}}{{Rupke}
  et~al.}{2005a}]{rupke05a}
{Rupke} D.~S.,  {Veilleux} S.,    {Sanders} D.~B.,  2005a, \apjs, 160, 87

\bibitem[\protect\citeauthoryear{{Rupke}, {Veilleux} \& {Sanders}}{{Rupke}
  et~al.}{2005b}]{rupke05b}
{Rupke} D.~S.,  {Veilleux} S.,    {Sanders} D.~B.,  2005b, \apjs, 160, 115

\bibitem[\protect\citeauthoryear{{Salim}, {Rich}, {Charlot}, {Brinchmann},
  {Johnson}, {Schiminovich}, {Seibert}, {Mallery}, {Heckman}, {Forster},
  {Friedman}, , {Small}, {Wyder}, {Bianchi}, {Donas}, {Lee} \& {Yi}}{{Salim}
  et~al.}{2007}]{salim07}
{Salim} S.,  {Rich} R.~M.,  {Charlot} S.,  {Brinchmann} J.,  {Johnson} B.~D.,
  {Schiminovich} D.,  {Seibert} M.,  {Mallery} R.,  et al.,  2007, \apjs, 173, 267

\bibitem[\protect\citeauthoryear{{S{\'a}nchez-Bl{\'a}zquez}, {Peletier},
  {Jim{\'e}nez-Vicente}, {Cardiel}, {Cenarro}, {Falc{\'o}n-Barroso}, {Gorgas},
  {Selam} \& {Vazdekis}}{{S{\'a}nchez-Bl{\'a}zquez} et~al.}{2006}]{sanchez06}
{S{\'a}nchez-Bl{\'a}zquez} P.,  {Peletier} R.~F.,  {Jim{\'e}nez-Vicente} J.,
  {Cardiel} N.,  {Cenarro} A.~J.,  {Falc{\'o}n-Barroso} J.,  {Gorgas} J.,
  {Selam} S.,    et al.,  2006, \mnras, 371, 703

\bibitem[\protect\citeauthoryear{{Schlegel}, {Finkbeiner} \&
  {Davis}}{{Schlegel} et~al.}{1998}]{schlegel98}
{Schlegel} D.~J.,  {Finkbeiner} D.~P.,    {Davis} M.,  1998, \apj, 500, 525

\bibitem[\protect\citeauthoryear{{Schwartz} \& {Martin}}{{Schwartz} \&
  {Martin}}{2004}]{schwartz04}
{Schwartz} C.~M.,  {Martin} C.~L.,  2004, \apj, 610, 201

\bibitem[\protect\citeauthoryear{{Shapiro} \& {Field}}{{Shapiro} \&
  {Field}}{1976}]{shapiro76}
{Shapiro} P.~R.,  {Field} G.~B.,  1976, \apj, 205, 762

\bibitem[\protect\citeauthoryear{{Shapley}, {Steidel}, {Pettini} \&
  {Adelberger}}{{Shapley} et~al.}{2003}]{shapley03}
{Shapley} A.~E.,  {Steidel} C.~C.,  {Pettini} M.,    {Adelberger} K.~L.,  2003,
  \apj, 588, 65

\bibitem[\protect\citeauthoryear{{Silk}}{{Silk}}{2003}]{silk03}
{Silk} J.,  2003, \mnras, 343, 249

\bibitem[\protect\citeauthoryear{{Silk} \& {Rees}}{{Silk} \&
  {Rees}}{1998}]{Silk98}
{Silk} J.,  {Rees} M.~J.,  1998, \aap, 331, L1

\bibitem[\protect\citeauthoryear{{Smith}, {Tucker}, {Kent}, {Richmond},
  {Fukugita}, {Ichikawa}, {Ichikawa} \& {Jorgensen}}{{Smith}
  et~al.}{2002}]{smith02}
{Smith} J.~A.,  {Tucker} D.~L.,  {Kent} S.,  {Richmond} M.~W.,  {Fukugita} M.,
  {Ichikawa} T.,  {Ichikawa} S.,    {Jorgensen} A.~M.,  et al., 2002, \aj, 123, 2121

\bibitem[\protect\citeauthoryear{{Strauss}, {Weinberg}, {Lupton}, {Narayanan},
  {Annis}, {Bernardi}, {Blanton} \& {Burles}}{{Strauss}
  et~al.}{2002}]{strauss02}
{Strauss} M.~A.,  {Weinberg} D.~H.,  {Lupton} R.~H.,  {Narayanan} V.~K.,
  {Annis} J.,  {Bernardi} M.,  {Blanton} M.,    {Burles} S.,  2002, \aj, 124,
  1810

\bibitem[\protect\citeauthoryear{{Strickland}, {Heckman}, {Colbert}, {Hoopes}
  \& {Weaver}}{{Strickland} et~al.}{2004}]{strickland04}
{Strickland} D.~K.,  {Heckman} T.~M.,  {Colbert} E.~J.~M.,  {Hoopes} C.~G.,
  {Weaver} K.~A.,  2004, \apj, 606, 829

\bibitem[\protect\citeauthoryear{{Stringer}, {Benson}, {Bundy}, {Ellis} \&
  {Quetin}}{{Stringer} et~al.}{2009}]{stringer08}
{Stringer} M.~J.,  {Benson} A.~J.,  {Bundy} K.,  {Ellis} R.~S.,    {Quetin}
  E.~L.,  2009, \mnras, 393, 1127

\bibitem[\protect\citeauthoryear{{Tremonti}, {Heckman}, {Kauffmann},
  {Brinchmann}, {Charlot}, {White}, {Seibert}, {Peng}, {Schlegel}, {Uomoto},
  {Fukugita} \& {Brinkmann}}{{Tremonti} et~al.}{2004}]{tremonti04}
{Tremonti} C.~A.,  {Heckman} T.~M.,  {Kauffmann} G.,  {Brinchmann} J.,
  {Charlot} S.,  {White} S.~D.~M.,  {Seibert} M.,  {Peng} E.~W.,  et al.,  2004, \apj, 613, 898

\bibitem[\protect\citeauthoryear{{Tremonti}, {Moustakas} \&
  {Diamond-Stanic}}{{Tremonti} et~al.}{2007}]{tremonti07}
{Tremonti} C.~A.,  {Moustakas} J.,    {Diamond-Stanic} A.~M.,  2007, \apjl,
  663, L77

\bibitem[\protect\citeauthoryear{{Veilleux}, {Cecil} \&
  {Bland-Hawthorn}}{{Veilleux} et~al.}{2005}]{veilleux05}
{Veilleux} S.,  {Cecil} G.,    {Bland-Hawthorn} J.,  2005, \araa, 43, 769

\bibitem[\protect\citeauthoryear{{Weiner}, {Coil}, {Prochaska}, {Newman},
  {Cooper}, {Bundy}, {Conselice}, {Dutton}, {Faber}, {Koo}, {Lotz}, {Rieke} \&
  {Rubin}}{{Weiner} et~al.}{2009}]{weiner09}
{Weiner} B.~J.,  {Coil} A.~L.,  {Prochaska} J.~X.,  {Newman} J.~A.,  {Cooper}
  M.~C.,  {Bundy} K.,  {Conselice} C.~J.,  {Dutton} A.~A.,  et al.,  2009, \apj,
  692, 187

\bibitem[\protect\citeauthoryear{{Worthey}, {Faber}, {Gonzalez} \&
  {Burstein}}{{Worthey} et~al.}{1994}]{worthey94}
{Worthey} G.,  {Faber} S.~M.,  {Gonzalez} J.~J.,    {Burstein} D.,  1994,
  \apjs, 94, 687

\bibitem[\protect\citeauthoryear{{York}, {Adelman}, {Anderson} Jr., {Anderson},
  {Annis}, {Bahcall}, {Bakken} \& {Barkhouser}}{{York} et~al.}{2000}]{york00}
{York} D.~G.,  {Adelman} J.,  {Anderson} Jr. J.~E.,  {Anderson} S.~F.,  {Annis}
  J.,  {Bahcall} N.~A.,  {Bakken} J.~A.,    {Barkhouser} R., et al., 2000, \aj, 120,
  1579

\end{thebibliography}

\begin{appendix}
Here we compare the stellar continuum of \citet[][hereafter GD05]{gonzalez05} 
and CB08 stellar population models to check whether our results are 
model dependent.

\begin{figure*}
\bc
\hspace{-1.6cm}
\resizebox{17cm}{!}{\includegraphics{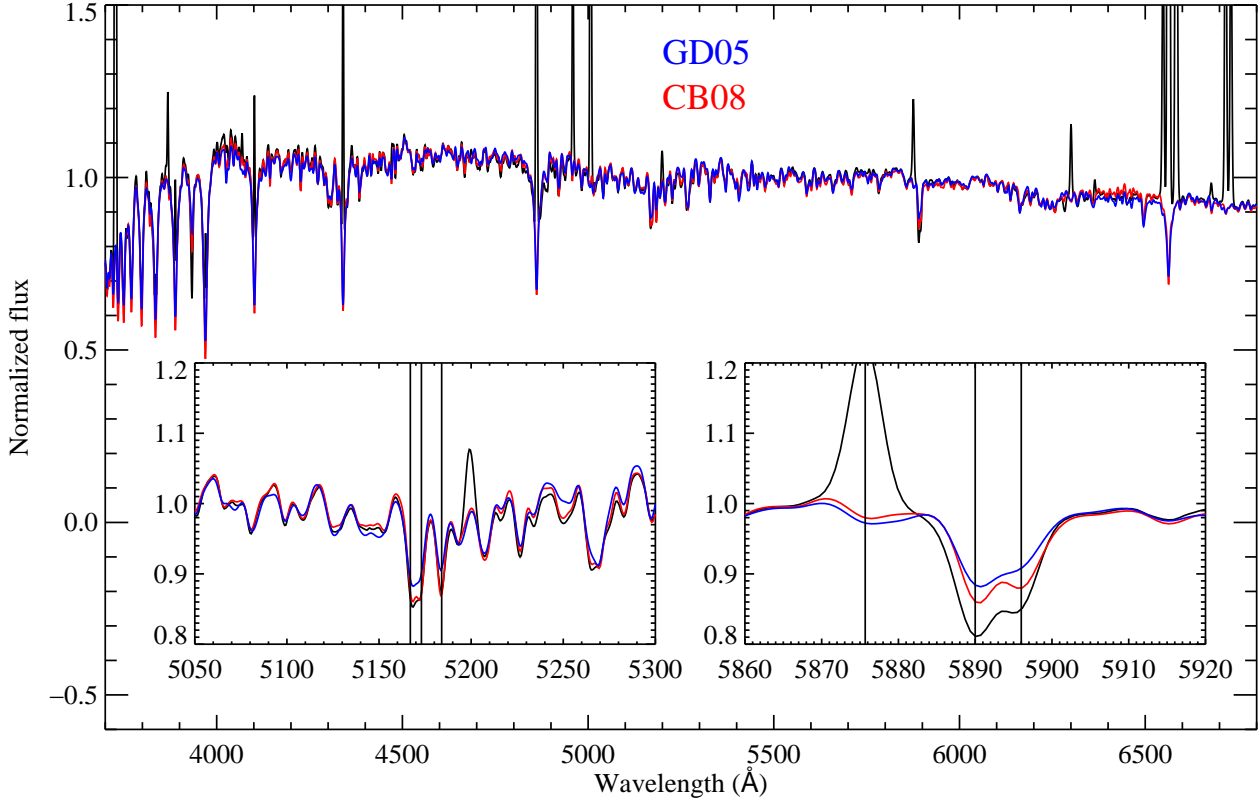}}\\%
\caption{This figure is the same as Figure~\ref{NaD_cont_example}, but we add 
the GD05 stellar continuum model (blue line) for comparison.  The 
CB08 model provides a better fit to the stellar population.
\label{cb08_gc05}}
\ec
\end{figure*}

In \S2.1, we noted that stellar Na~D appears to be over-estimated in our
continuum fits, resulting in emission at the systemic velocity in the
continuum-normalized spectra.  The libraries of stellar spectra used
by CB08 are empirical, so Na D could be contaminated by Milky Way
absorption.  One important question is to what extent our results are
affected by the stellar population models used to estimate the
continuum. In order to answer this question, we repeat all our
analysis using the models of GD05 which use fully theoretical stellar
libraries. Figure~\ref{cb08_gc05} shows both CB08 and GB05 model fits
to the stacked spectrum shown in Figure~\ref{NaD_cont_example}.  The CB08
model is shown in red and the GD05 model in blue.  The CB08
model clearly provides a better fit of the the full spectrum and the
region around the Mg I triplet.

Using the GD05 theoretical models, all the trends shown in
Figs. \ref{Fig_NaD_inc_param}, \ref{Fig_NaD_sys_phys}, and
\ref{Fig_NaD_dyn_phys} remain.  On average, the outflow velocity
decreases by $\sim 40 - 60$ $\rm km~s^{-1}$; the Na~D EW increases by
$\sim0.3$~\AA, for both disk-like and outflow components; and the
covering factors for the two components increase by 10\%. The line
widths for the systematic component are the same as that from CB08,
since they are bound to He I. The width of outflow component increases
by $\sim25 - 30$ $\rm km~s^{-1}$ and more scatter is introduced in the
trends with galaxy physical properties.  The largest change caused by
the using different stellar population models is the optical depth
$\tau_0$.  However, this parameter is not very strongly constrained in
either case.  In summary, the CB08 models provide a better overall fit
to the stellar continuum. However, our basic results are not sensitive
to the stellar population models used.
\end{appendix}

\end{document}